\documentclass[pra,twocolumn,superscriptaddress,10pt]{revtex4-1}
\usepackage{graphicx}
\usepackage{dcolumn}
\usepackage{color}
\usepackage{times}
\usepackage{bm}
\usepackage{amssymb}
\usepackage{amsmath}
\usepackage{epsfig}
\usepackage{epstopdf}
\usepackage{dsfont}
\usepackage{subfigure}
\usepackage{float}
\usepackage[colorlinks=true,citecolor=blue, linkcolor=blue, urlcolor=blue]{hyperref}
\begin{document}
\renewcommand{\thefootnote}{\fnsymbol {footnote}}
	
	\title{{Quantifying Quantumness in (A)dS spacetimes with Unruh-DeWitt Detector}}
	
	\author{Li-Juan Li}
	\affiliation{School of Physics and Optoelectronic Engineering, Anhui University, Hefei 230601, China}

   \author{Xue-Ke  Song}
\affiliation{School of Physics and Optoelectronic Engineering, Anhui University, Hefei 230601, China}

	\author{Liu Ye}
	%\email{yeliu@ahu.edu.cn}
	\affiliation{School of Physics and Optoelectronic Engineering, Anhui University, Hefei 230601, China}
%\affiliation{Center for Quantum Information, IIIS, Tsinghua University, Beijing 100084, People's Republic of China}
	\author{Dong Wang}
	\email{dwang@ahu.edu.cn}
    \affiliation{School of Physics and Optoelectronic Engineering, Anhui University, Hefei 230601, China}

\begin{abstract}{
{{Probing quantumness in curved spacetime is regarded as  one of fundamental and important topics in the framework of relativistic quantum information. In this work, we focus on the theoretical feasibility of probing quantum properties in  de Sitter (dS) and Anti-de Sitter (AdS) spacetimes via detectors. By employing the Unruh-DeWitt detector coupled with a massless scalar field, which is treated as an open system,  quantum uncertainty and quantum coherence in both dS and AdS spacetimes are investigated. Our analysis reveals that the acceleration in dS spacetime and the boundary conditions in AdS spacetime significantly impact the detector's evolution in the initial stage. Notably, both of the uncertainty and coherence will oscillate with the initial state being in a superposition state, however the high temperature
is able to suppress their oscillation. Interestingly, it is found that the constant values of the final uncertainty and coherence are identical as those in dS and AdS spacetimes, which are   determined by the ratio of energy gap to temperature. Hence, the current exploration offers insight into quantumness in dS and AdS   spacetimes,  and might be helpful to facilitate the curved-spacetime-based quantum information processing.}
 \vskip .2cm
 }}
\end{abstract}
\date{\today}
\maketitle

\section{Introduction} %第1节 引言
{The combination of relativity and quantum information has given rise to a new and promising discipline: relativistic quantum information \cite{T13}. By introducing relativistic effects,  quantum phenomena under extreme spacetime conditions can be explored, especially in curved spacetime. De Sitter (dS) spacetime, Anti-de Sitter (AdS) spacetime, and flat Minkowski spacetime,  as solutions to Einstein's field equations of general relativity, are known as three types of maximally symmetric constant curved spacetimes, which are typical and significant in Physics. Among them, de Sitter spacetime describes a universe with positive curvature, playing a crucial role in understanding quantum gravity and inflationary cosmology. Anti-de Sitter spacetime, on the other hand, has negative curvature and is often used as a background in string theory, providing important theoretical support for the study of quantum gravity and black hole physics.

The Unruh-DeWitt (UDW) detectors \cite{55,56} are widely used to investigate the interaction between quantum fields and detectors in different curved spacetime backgrounds. The most famous discovery is the Unruh effect \cite{55}: if the detector is coupled to a quantum field in the Minkowski vacuum and moves along a constant acceleration trajectory, the detector will detect a spectrum of particles with thermal Planckian shape and the temperature proportional to the acceleration of the detector.}

{Therefore, to investigate quantum characteristics in dS and AdS spacetimes, the UDW detector coupled to a massless scalar field can been considered, which is usually seen} as an open quantum system \cite{53,54}. Explicitly, the Unruh-DeWitt can be seen as the system, while the quantum field of the noisy vacuum can be seen as the environment. This detector is essentially a two-level quantum system with energy gap $\Omega$ between ${\left| 0 \right\rangle _D}$ and ${\left| 1 \right\rangle _D}$. Consider the total Hamiltonian of a combined system containing detector and quantum field
\begin{align}
H_{tot}=H_{D}+H_{\phi}+H_{I},
\label{t1}
\end{align}
where ${H_\phi } = \sum\limits_k {{\omega _k}a_k^\dag {a_k}} $ is the Hamiltonian of the massless scalar field $\phi(x)$, and the free Hamiltonian of the detector is expressed as
\begin{align}
H_{D}&=\frac{1}{2} \Omega a_{D}^{\dagger} a_{D} \nonumber\\
&=\frac{1}{2}\Omega ({\left| 0 \right\rangle _D}{\left\langle 0 \right|_D} - {\left| 1 \right\rangle _D}{\left\langle 1 \right|_D}).
\end{align}
{And the} Hamiltonian of the interaction between quantum field and detector $H_{I}$ can be written as \cite{T9}
\begin{align}
H_{I}=\lambda\left(e^{i \Omega \tau} \sigma^{+}+e^{-i \Omega \tau} \sigma^{-}\right) \otimes \phi[x(\tau)],
\label{t2}
\end{align}
where $\lambda$ is a constant that represents the strength of the interaction, $\sigma^{+}={\left| 1 \right\rangle _D}{\left\langle 0 \right|_D}$ and $\sigma^{-}{\left| 0 \right\rangle _D}{\left\langle 1 \right|_D}$ are ladder operators on the Hilbert space associated with the detector. {The Unruh-DeWitt detector moves along a trajectory $x(\tau)$, parametrized by the proper time $\tau$. The trajectory plays a central role in the detector's interaction with the quantum field, as it determines how the detector interacts with the quantum field and its response to it. The different trajectories, like stationary trajectories or acceleration trajectories, will affect the response of the detector and the detection result.  From the response of the detector, we can infer the properties of the field in a region of spacetime. This can be demonstrated explicitly by studying the response of an Unruh-DeWitt detector coupled to a quantum field in different states of motion.} {The applications of UDW detectors in curved spacetime have undergone continuous evolution \cite{T9,58,59,T10,T11,T12}. The UDW detector can distinguish the structure of the quantum vacuum. In AdS spacetime, if the acceleration is sufficiently large, the UDW detector will detect thermal radiation  \cite{59}. Furthermore, the detector can also be utilized as a probe to calculate the Fisher information in curved spacetime \cite{T9,T12}.}

{ The uncertainty relation \cite{1} and quantum coherence \cite{42} play a crucial role in quantum information science.} The uncertainty principle is regarded as one of the dramatic criteria that quantum mechanics is distinguished from classical counterpart. {Quantum coherence arises from the superposition of quantum states and is considered one of the essential quantum resources in quantum information processing. However, it  often suffers from environmental noises, resulting in the system's decoherence. Therefore, to clarify the measured uncertainty and quantum coherence is indispensable and
play a vital role in practical quantum computing and quantum communication \cite{9,10,11,12,13,14,15,T1,T2,T3,T4,T5,T6,T7,T14}.}

{In the decades}, significant advances have emerged in exploring quantum information theory within the framework of quantum field theory across both flat and curved spacetimes.
{Recently}, many scientists have made significant progress in the field of relativistic quantum information science in Minkowski spacetime \cite{45,46,47,48,49,50,51}.
{However, quantum fields behave differently in curved spacetime compared to flat spacetime, with dS and AdS playing crucial roles in Quantum Gravity Theory. Therefore, understanding the quantum properties of these spacetimes is essential for advancing the development of quantum gravity theory. Additionally, since dS and AdS spacetimes involve important thermodynamic characteristics, investigating the quantum uncertainty and coherence of quantum fields in these spaces helps uncover the deep connection between quantum gravity and thermodynamics. Furthermore, in curved spacetime, quantum fields may undergo decoherence due to expansion or gravitational effects. Examining how quantum coherence evolves in curved spacetimes provides valuable insights into how spacetime affects the storage and processing of quantum information. Motivated by these considerations, this article aims to investigate the quantum properties of dS and AdS spacetimes by using the UDW detector.}

The structure of this article is as follows. {In Section II, we introduce the uncertainty relations, quantum coherence, dS and AdS spacetimes. Furthermore, we also introduce the corresponding response functions of the  inertial UDW detector model in the two spacetimes.}
In Section III, we provide the solution of  the master equation of the UDW detector.
The uncertainty and quantum coherence in dS and AdS spacetimes are elucidated in Sections IV and V, respectively. Finally, we end our paper with a brief conclusion.

\section{preliminary}
\subsection{{Uncertainty Relations and Quantum Coherence}}
 In 1927, Heisenberg \cite{1} first proposed the uncertainty principle, which demonstrates that it is impossible to  accurately predict the outcomes to measure a pair of non-commuting observables, i.e., momentum and position. Subsequently, Kennard \cite{2} and Robertson \cite{3} further proposed the new inequality
with respect to an arbitrary pair of incompatible measurements
\begin{align}
\Delta \hat{Q}\cdot\Delta \hat{R} \ge \frac{1}{2}| {\langle {[\hat{Q},\hat{R}]}\rangle } |,
\label{f1}
\end{align}
where the standard deviation $\Delta  \hat{Q}  = {\rm{ }}\sqrt {\left\langle {{ \ \hat{Q} ^2}} \right\rangle  - {{\left\langle  \hat{Q}  \right\rangle }^2}} $ is the standard deviation with $\langle \hat{Q} \rangle  = {\rm Tr}(\rho \hat{Q})$ being the expected value, and $[ {\hat{Q},\hat{R}} ] = \hat{Q}\hat{R} - \hat{R}\hat{Q}$ represents the commutator associated with the
observables $\hat{Q}$ and $\hat{R}$. However, due to that the lower bound of (\ref{f1}), there exists a shortcoming for Robertson's inequality, the lower bound  of the inequality will become zero as the prepared state  is one of the eigenstates of $\hat{Q}$ and $\hat{R}$, implying a trivial result in this case.
To overcome this defect,  Deutsch \cite{4}  proposed a new uncertainty relation based on Shannon entropy. Then this relation was further optimized by Kraus \cite{5} and Maassen and Uffink \cite{6}, which is expressed as
\begin{align}
H( \hat{Q} ) + H( \hat{R} ) \ge \log_2 \frac{1}{{c( {\hat{Q},\hat{R}} )}} = :{q_{MU}},
\label{f2}
\end{align}
where $H( \hat{Q }) =  - \sum\limits_i {{p_i}\log_2 {p_i}} $ is Shannon entropy of operator $\hat{Q }$, and ${p_i} = \langle {q_i}|\rho |{q_i}\rangle $ is the probability distribution of the $i$-th outcome with the eigenstates $|{q_i}\rangle$ of $\hat{Q}$, the overlap $c( {\hat{Q},\hat{R}} ) =\max _{ij} {| {\langle {q _i|r _j} \rangle } |^2}$, $| {q _i} \rangle$ and  $| {r _j} \rangle$ denote the eigenvectors of $\hat{Q}$ and $\hat{R}$, respectively.

{It is worth noting that} the uncertainty relations mentioned above are applicable to single-particle systems, a natural question is raised: how to express the relation when the measured particle is correlated with another? Up to 2010, Berta {\it et al.} \cite{7} had proposed a novel form of bipartite uncertainty relation, so-called quantum-memory-assisted entropic uncertainty relation (QMA-EUR), which can be expressed mathematically as
\begin{align}
S(\hat{Q}|B) + S(\hat{R}|B) \ge \log_2 \frac{1}{{c( {\hat{Q},\hat{R}} )}} +S(A|B),
\label{f3}
\end{align}
where $S(\hat{Q}|B) = S({\rho _{\hat{Q}B}}) - S({\rho _B})$  represents the conditional von Neumman entropy \cite{8} of the post-measurement state ${\rho _{\hat{Q}B}} = \sum\limits_i {(|{q_i}\rangle \langle {q_i}| \otimes {\mathds{I}_B}){\rho _{AB}}(|{q_i}\rangle \langle {q_i}|\otimes {\mathds{I}_B})}$ with $S\left( \rho  \right) =  - {\rm tr}\left( {\rho \log_2 \rho } \right)$, and $\mathds{I}_B$ is an identical operator in the Hilbert space of $B$.
$S(A|B) = S({\rho _{AB}}) - S({\rho _B})$ denotes the conditional entropy of systematic density operator $\rho _{AB}$.
Interestingly, this relation can be interpreted through the uncertainty game scenario: Bob prepares two entangled particles $A$ and $B$, and sends $A$ to Alice, then keeps $B$ as a quantum memory.
Alice selects one operator to measure $A$ and informs Bob of her choice. Remarkably, Bob can predict measurement outcomes with minimal deviation, as constrained by this relation.

As a matter of fact, the QMA-EUR can be widely used in many fields of quantum information and communication, including entanglement witnessing \cite{9,10}, quantum key distribution  \cite{11,12,13}, EPR steering \cite{14,15}, wave-particle duality \cite{T1,T2}, quantum metrology \cite{T3}, quantum randomness \cite{T4}, quantum teleportation \cite{T5}, and quantum cryptography \cite{T6,T7}. Moreover, it is generally applicable to probe the quantumness of diverse systems \cite{16,17}, for examples, neutrino systems \cite{18,19,20}, the Heisenberg spin-chain model \cite{21,23} and the curved spacetimes \cite{24,25,26,27,T8}. To date, considerable efforts have been dedicated to improving bipartite and multipartite QMA-EUR \cite{28,29,30,31,32,33,34,35,36,37,38,39,40,41}.

In addition, quantum coherence \cite{42} originated from the principle of superposition of states is also regarded as a key quantum resource that can be widely used for implementing various quantum tasks. Various methods were presented to quantify coherence \cite{43,44}. Specifically, the $l_1$-norm is particularly prevalent in quantum physics, which is defined as follows:
\begin{align}
C(\rho ) =\sum\limits_{i \ne j} {\left| {{\rho _{i,j}}} \right|},
\label{f4}
\end{align}
where $\rho _{i,j}$ represent the elements of systemic state $\rho$.
\subsection{{de Sitter and Anti-de Sitter spacetimes}}
Both dS$_4$ and AdS$_4$ can be represented as 4-dimensional hyperboloids embedded in 5-dimensional spacetime
\begin{align}
d{s^2} =  - dX_0^2 + dX_1^2 + dX_2^2 + dX_3^2 \pm dX_4^2.
\end{align}
(A)dS$_4$ can be described as the hyperboloid
\begin{align}
X_0^2 + X_1^2 + X_2^2 + X_3^2 \pm X_4^2 =  \pm {l^2},
\end{align}
where the quantity $l$ is referred to the (A)dS length, and the plus and minus sign represent dS$_4$ and AdS$_4$, respectively. The static metric is obtained from the flat space embedding via the transformations
\begin{align}
\begin{array}{*{20}{l}}
\begin{array}{l}
{X_1} = r\sin \theta \sin \phi,\ {X_2} = r\sin \theta \cos \phi,\ {X_3} = r\cos \theta ,\\
{X_0} = \left\{ {\begin{array}{*{20}{l}}
{\sqrt {{l^2} - {r^2}} \sinh (t/l),\;{\rm{for}}\;{\rm{dS}}\;{\rm{spacetime}},}\\
{\sqrt {{l^2} + {r^2}} \cosh (t/l),\;{\rm{for}}\;{\rm{AdS}}\;{\rm{spacetime}},}
\end{array}} \right.
\end{array}\\
{X_4} = \left\{ {\begin{array}{*{20}{l}}
{\sqrt {{l^2} - {r^2}} \sin (t/l),\;{\rm{for}}\;{\rm{dS}}\;{\rm{spacetime}},}\\
{\sqrt {{l^2} + {r^2}} \cos (t/l),\;{\rm{for}}\;{\rm{AdS}}\;{\rm{spacetime}}.}
\end{array}} \right.
\end{array}
\end{align}
Thus, the static metric for (A)dS can be obtained as
\begin{align}
d{s^2} =  - (1 \pm \frac{{{r^2}}}{{{l^2}}})d{t^2} + {(1 \pm \frac{{{r^2}}}{{{l^2}}})^{ - 1}}d{r^2} + {r^2}d\Omega _2^2.
\label{t3}
\end{align}
In addition to the static metric, we also consider the comoving coordinates for dS$_4$, the corresponding transformations
\begin{align}
\left\{ {\begin{array}{*{20}{l}}
{T = l\sinh (t/l) + \frac{{{r^2}{e^{t/l}}}}{{2l}}},\\
{{X_0} = l\cosh (t/l) - \frac{{{r^2}{e^{t/l}}}}{{2l}}},\\
{{X_1} = {e^{t/l}}{x_1}},\\
{{X_2} = {e^{t/l}}{x_2}},\\
{{X_3} = {e^{t/l}}{x_3}},
\end{array}} \right.
\end{align}
then we end up with the comoving metric
\begin{align}
d{s^2} =  - d{t^2} + {e^{2t/l}}(dx_1^2 + dx_2^2 + dx_3^2).
\label{t4}
\end{align}
{The massless scalar field $\phi$ conformally coupled to curvature via the action
\begin{align}
S = \int {\sqrt { - g} [\frac{1}{2}{g^{\mu \nu }}{\nabla _\mu }\phi {\nabla _\nu }\phi  - \frac{1}{{12}}R{\phi ^2}]}.
\end{align}}
The Wightman function of scalar field is $W(x,x')=\left\langle 0 \right|\phi (x)\phi (x')\left| 0 \right\rangle $, and for a comoving observer in the conformal vacuum of dS$_4$, the Wightman function is known as
\begin{align}
{W^{dS}}(x(\tau ),x(\tau ')) =  - \frac{1}{{2\sqrt 2 \pi l}}[\frac{1}{{{{\sinh }^2}(\frac{{\tau  - \tau '}}{l} - i\varepsilon )}}],
\label{t5}
\end{align}
here, the comoving trajectory is stationary, because this is only a function of $\Delta \tau= \tau  - \tau '$. {The conformal vacuum obeys the symmetry of the de Sitter group \cite{61}.}

Likewise, for AdS, the Wightman function \cite{57} can be expressed as
\begin{align}
{W^{AdS}}(x,x') =  - \frac{1}{{4\sqrt 2 \pi l}}[\frac{1}{{\sigma (x,x')}} - \frac{\zeta }{{\sigma (x,x') + 2}}],
\label{t6}
\end{align}
where $2l^2\sigma (x,x')$ means the geodesic distance between $x$ and $x'$, and $\sigma (x,x')$ can be computed from the embedding space. The different $\zeta$ represent different boundary conditions: Dirichlet ($\zeta= 1$), transparent ($\zeta= 0$), and Neumann ($\zeta=-1$).

Let's go back to the detector, for the detector, the probability of jumping from the ground state ${\left| 0 \right\rangle _D}$ to the excited state ${\left| 1 \right\rangle _D}$ is proportional to the response function \cite{58,59}, which can be expressed as
\begin{align}
G(\omega ) = \int_{ - \infty }^\infty  {d\tau } \int_{ - \infty }^\infty  {d\tau '{e^{-i\omega (\tau  - \tau ')}}{W}(x(\tau ),x(\tau '))}.
\end{align}
{The difference in trajectory will affect the interaction between the detector and the quantum field, and then change the response function of the detector.}
As mentioned above, the comoving trajectory is stationary. As a result, the response function can be further expressed as the response per unit time
\begin{align}
G(\omega ) = \int_{ - \infty }^\infty  {d\Delta \tau } {e^{ - i\omega \Delta \tau }}W(\Delta \tau ).
\label{t7}
\end{align}

In this paper, we mainly examine the quantumness of the UDW detector in the following three cases: (1) a
comoving detector in dS$_4$ spacetime, (2) a uniformly accelerated detector in dS$_4$ spacetime, and (3) a uniformly accelerated detector in AdS$_4$ spacetime.

For comoving detector in dS$_4$ spacetime, the comoving observer experiences a thermal bath of radiation which is called the Gibbons-Hawking effect, and the temperature $T = \frac{1}{{2\pi l}}$ called the Gibbons-Hawking temperature \cite{60}. {For simplification, Planck units $G = c = \hbar  = {k_B} = 1$ are considered here.} Thereby, the response function per unit time in terms of $T$ is \cite{61}
\begin{align}
{G^{dS}}(\omega ) = \frac{\omega }{{2\pi }}\frac{1}{{{e^{\omega /T}} - 1}},
\label{t8}
\end{align}
which is obtained by Fourier transform of the Wightman function (\ref{t5}) with contour integration.

With respect to the accelerated detector in dS$_4$, the detector suffers from both the Unruh effect and the Gibbons-Hawking effect, so the temperature $T = \frac{{\sqrt {{a^2} + 1/{l^2}} }}{{2\pi }}$ \cite{62,63}. According to Ref. \cite{64} and Eq. (\ref{t4}), one can get the response per unit time
\begin{align}
G_{accel}^{dS}(\omega ) = \frac{\omega }{{4{\pi ^2}}}\frac{1}{{lT}}\frac{1}{{{e^{\omega /T}} - 1}}.
\label{t9}
\end{align}
When $a \to 0$ and $T \to 1/2\pi l$, Eq. (\ref{t9}) is equal to Eq. (\ref{t8}).

A uniformly accelerated detector in AdS$_4$ spacetime also experiences a nonzero temperature $T = \frac{{\sqrt {{a^2}{l^2} - 1} }}{{2\pi l}}$ \cite{62}, the response function \cite{59} can be obtained as

\begin{widetext}
\begin{align}
G_{accel}^{AdS}(\omega ) = \left\{ {\frac{\omega }{{2\pi }} - \frac{\zeta }{{4\pi l\sqrt {4{\pi ^2}{T^2}{l^2} + 1} }}\sin[\frac{\omega }{{\pi T}}\sinh^{ - 1}(2\pi Tl)]} \right\}\frac{{\Theta (T)}}{{{e^{\omega /T}} - 1}},
\label{t10}
\end{align}
\end{widetext}
where $\Theta (*)$ is the Heaviside function. If  $\zeta= 0$, Eq. (\ref{t10}) is equivalent to the response function (\ref{t8}) of comoving detector in dS spacetime.

%\begin{widetext}
%\begin{align}
%{G^{AdS}}(\omega ) = (\frac{\omega }{{2\pi }} - \frac{\zeta }{{4\pi a{l^2}}}\sin[\frac{{2\omega l}}{{{a^2}{l^2} - 1}}\sinh^{ - 1}(\sqrt {{a^2}{l^2} - 1} )])\frac{{\Theta (al - 1)}}{{\exp (\frac{{2\pi \omega l}}{{\sqrt {{a^2}{l^2} - 1} }}) - 1}}
%\end{align}
%\end{widetext}

\section{Master equation}

\begin{figure*}
\centering
{
\includegraphics[width=6.5cm]{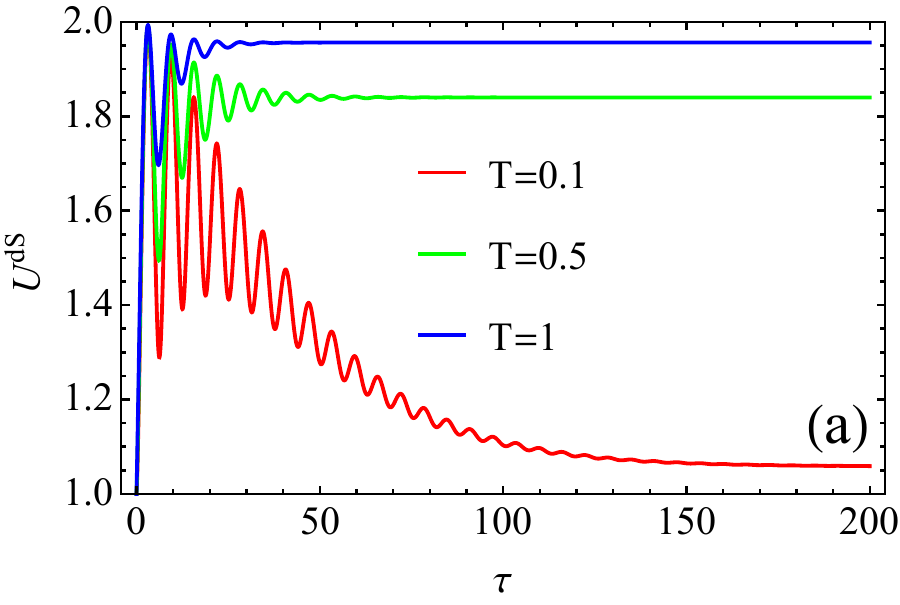}} \ \ \ \
{
\includegraphics[width=6.5cm]{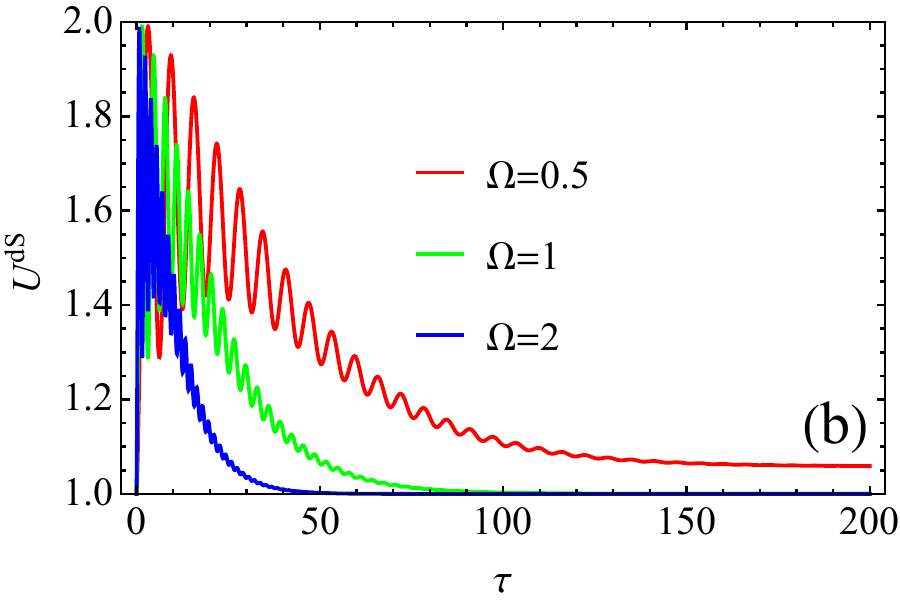}}\ \ \ \
{
\includegraphics[width=6.5cm]{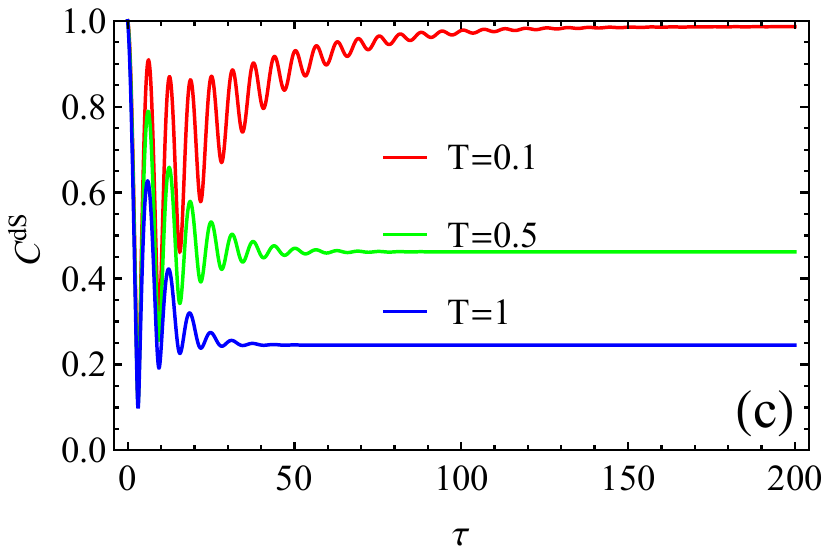}} \ \ \ \
{
\includegraphics[width=6.5cm]{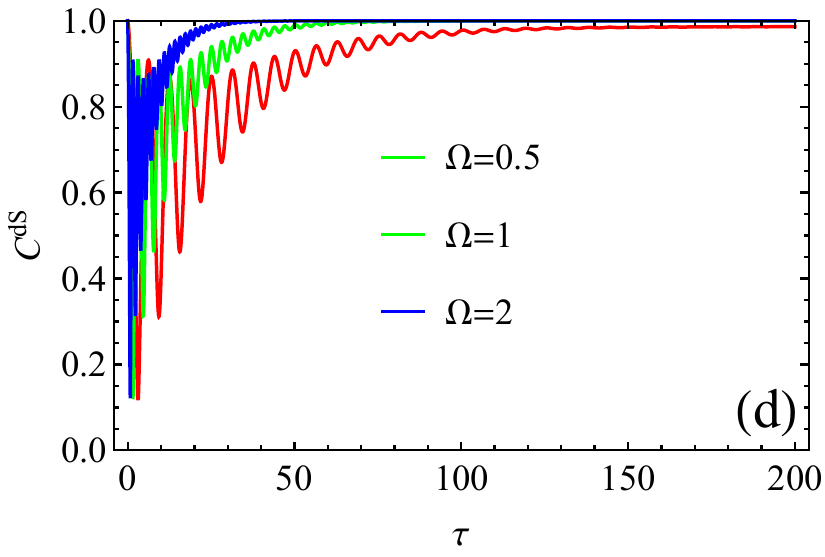}}
\caption{The time evolution of measurement uncertainty and coherence probed by comoving detector in dS spacetime. (a) The time evolution of the measurement uncertainty $U^{dS}$ with different Gibbons-Hawking temperature $T$ and fixed energy gap $\Omega=0.5$ and state parameter $\theta=\frac{\pi}{2}$. (b) The time evolution of the measurement uncertainty $U^{dS}$ with different energy gap $\Omega$ and fixed Gibbons-Hawking temperature $T=0.1$ and state parameter $\theta=\frac{\pi}{2}$. (c) The time evolution of coherence $C^{dS}$  with different Gibbons-Hawking temperature $T$ and  fixed energy gap $\Omega=0.5$ and state parameter $\theta=\frac{\pi }{2}$. (d)The time evolution of the coherence $C^{dS}$ with different energy gap $\Omega$ and fixed Gibbons-Hawking temperature $T=0.1$ and state parameter $\theta=\frac{\pi}{2}$.}
\label{fig.1}
\end{figure*}
The time evolution of combined system $\rho _{tot}$ can be described by the von Neumann equation
\begin{align}
\frac{{\partial {\rho _{tot}}}}{{\partial \tau }} =  - i[H,{\rho _{tot}}],
\end{align}
where the initial state ${\rho _{tot}}(0) = {\rho _D}(0) \otimes \left| 0 \right\rangle \left\langle 0 \right|$, and ${\rho _D}(0)$ is the initial state of the detector and $\left| 0 \right\rangle$ is the conformal vacuum of the scalar field $\phi(x)$. We can trace over the field of the system to attain the state of the detector, ${\rho _D} = {\rm tr}{_\phi }{\rho _{tot}}$. For the weak coupling and Markovian approximation between the detector and the field, the evolution of the detector can be described by the Kossakowski-Lindblad master equation \cite{54}
\begin{align}\label{s1}
\frac{{\partial {\rho _D}(\tau )}}{{\partial \tau }} =  - i[{H_{eff,}}{\rho _D}(\tau )] + L[{\rho _D}(\tau )],
\end{align}
where
\begin{align}
L[\rho ] = \frac{1}{2}\sum\limits_{i,j = 1}^3 {{C_{ij}}(2{\sigma _j}\rho {\sigma _i} - {\sigma _i}{\sigma _j}\rho  - \rho {\sigma _i}{\sigma _j})}.
\end{align}
Here, $\sigma _i$ are the Pauli matrices. The Kossakowski matrix $C_{ij}$ can be explicitly resolved as
\begin{align}
{C_{ij}} = \left( {\begin{array}{*{20}{c}}
A'&{ - iB'}&0\\
{iB'}&A'&0\\
0&0&{A' + C}
\end{array}} \right),
\end{align}
where
\begin{align}
\begin{array}{l}
A' = \frac{1}{2}[G(\Omega ) + G( - \Omega )],\\
B' = \frac{1}{2}[G(\Omega ) - G( - \Omega )],\\
C = G(0) - A'.
\end{array}
\end{align}

Considering the effect of the scalar field on the detector, the effective Hamiltonian of the detector is expressed as
\begin{align}
{H_{eff}} = \frac{1}{2}\tilde \Omega {\sigma _z},
\end{align}
$\tilde \Omega$ is a renormalized gap given by $\tilde \Omega  = \Omega  + i[K( - \Omega ) - K(\Omega )]$ with $K(\Omega ) = \frac{1}{{i\pi }}PV\int_{ - \infty }^\infty  {d\omega \frac{{G(\omega )}}{{\omega  - \Omega }}}$ being Hilbert transform of Wightman functions.

Assuming a general initial state $|\psi\rangle  = \sin \frac{\theta }{2}\left| 0 \right\rangle  + \cos \frac{\theta }{2}\left| 1 \right\rangle $, the master equation (\ref{s1}) can be analytically resolved, therefore, the density matrix $\rho (\tau )$ evolving over time can be expressed as
\begin{align}
\rho (\tau ) = \frac{1}{2}(\mathds{I} +\vec n(\tau ) \cdot \vec \sigma  ),
\end{align}
where $\vec \sigma  = ({\sigma _1},{\sigma _2},{\sigma _3})$ are Pauli matrices, $\mathds{I}$ denotes identity matrix, and the Bloch vector $ \vec n = ({n_1},{n_2},{n_3})$ reads
\begin{align}
\begin{array}{l}
{n_1}(\tau ) = {e^{ - A'\tau /2}}\sin \theta \cos \tilde \Omega \tau, \\
{n_2}(\tau ) = {e^{ - A'\tau /2}}\sin \theta \sin \tilde \Omega \tau, \\
{n_3}(\tau ) = {e^{ - A'\tau }}\cos \theta  - R(1 - {e^{ - A'\tau }}),
\end{array}
\end{align}
where $R = \frac{B'}{A'}$.

\section{Measurement Uncertainty and Coherence in de Sitter Spacetime}
\begin{figure*}
\centering
{
\includegraphics[width=5cm]{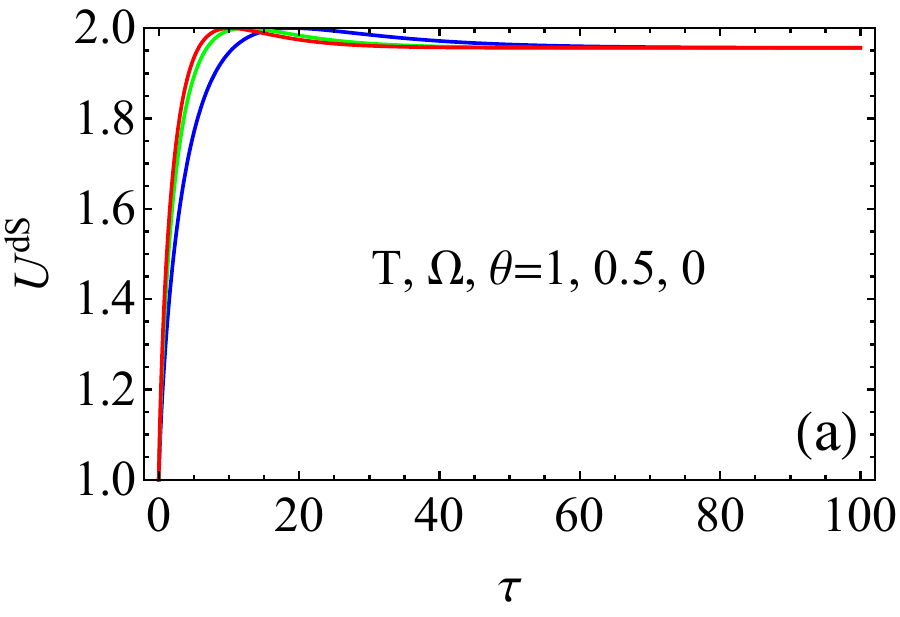}} \ \ \ \
{
\includegraphics[width=5cm]{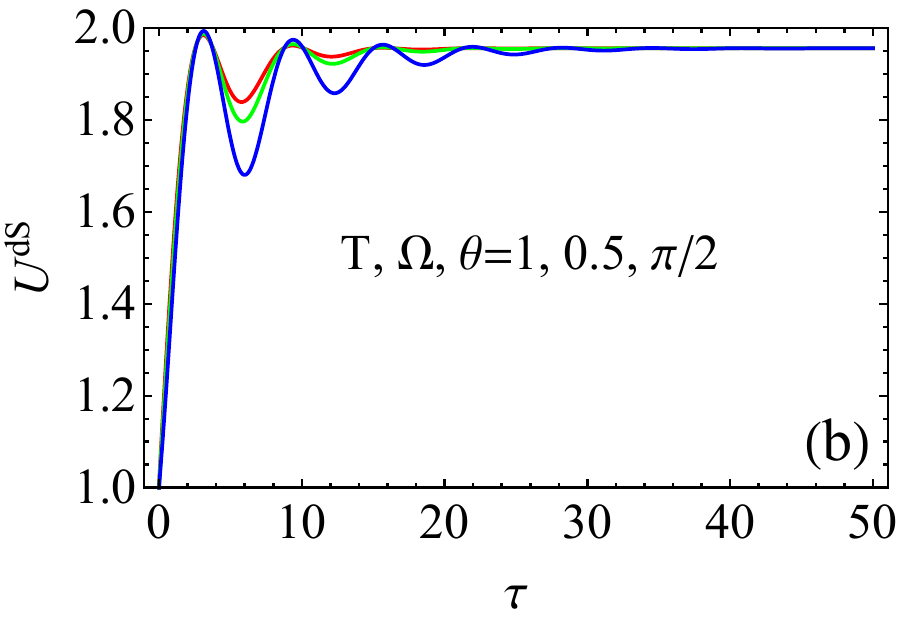}} \ \ \ \
{
\includegraphics[width=5cm]{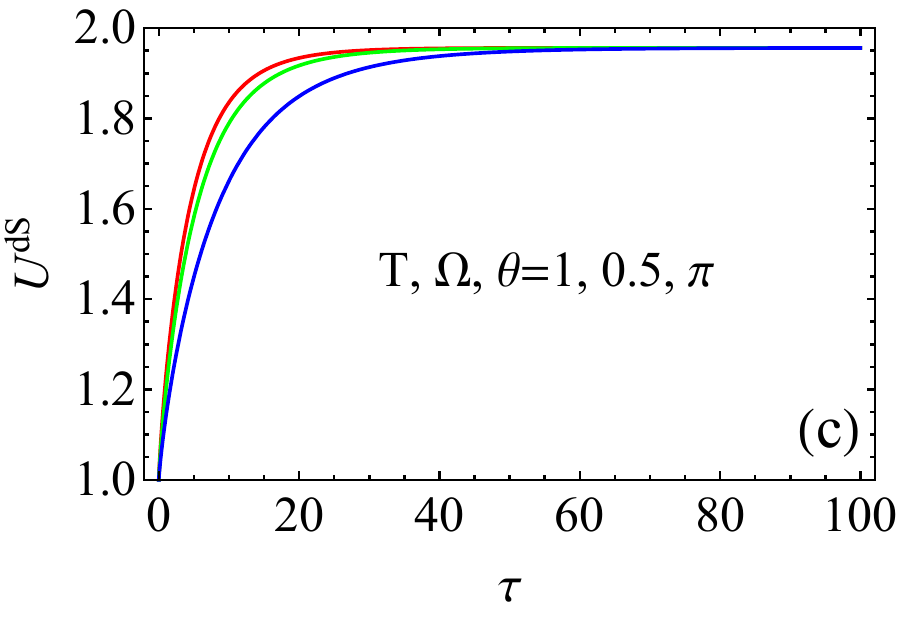}} \ \ \ \
{
\includegraphics[width=5cm]{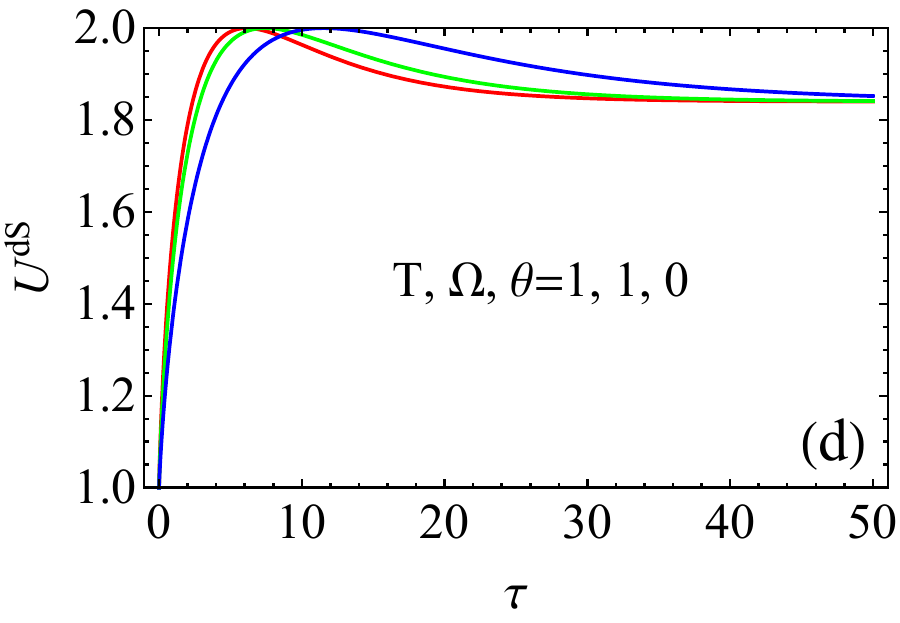}} \ \ \ \
{
\includegraphics[width=5cm]{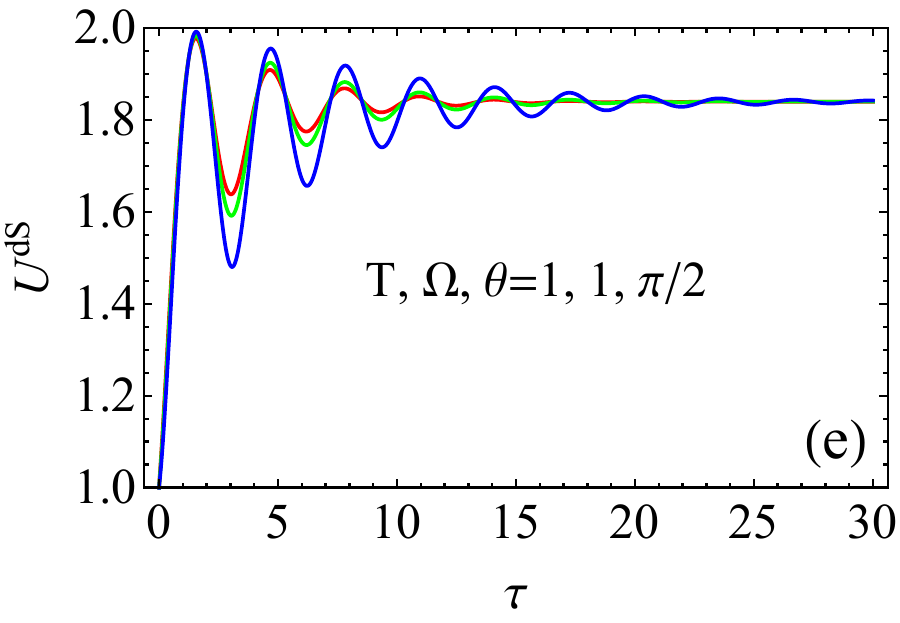}}\ \ \ \
{
\includegraphics[width=5cm]{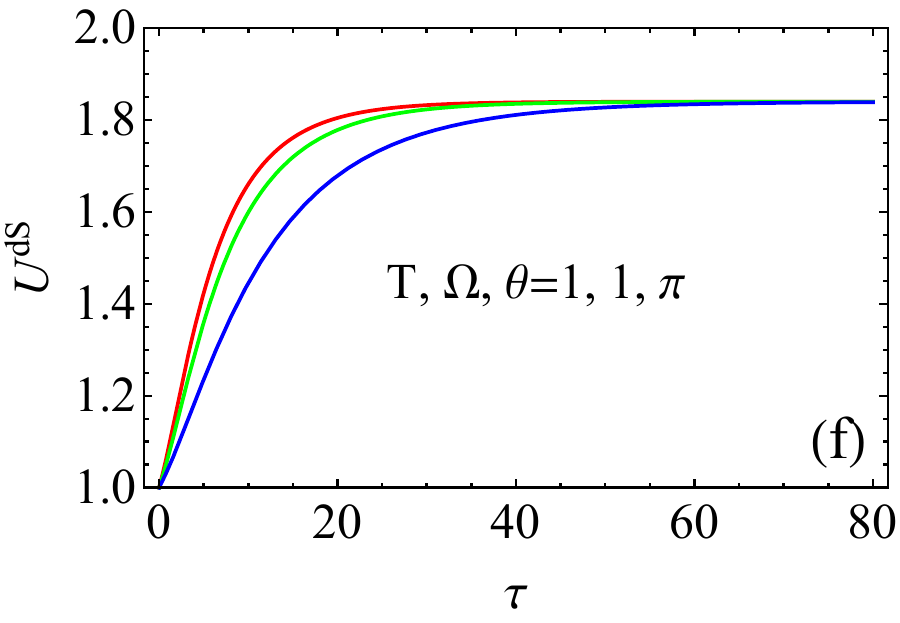}} \ \ \ \
{
\includegraphics[width=5cm]{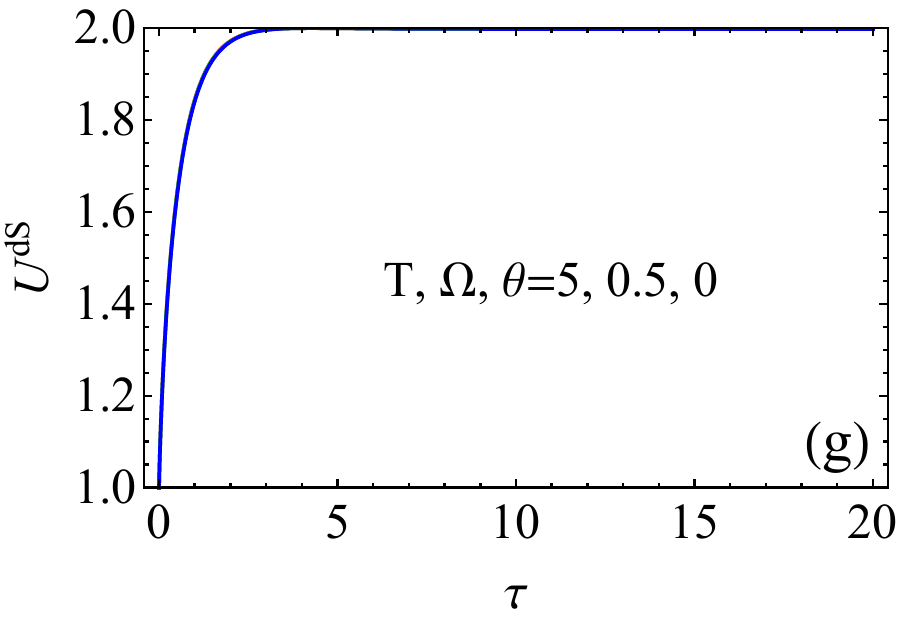}}\ \ \ \
{
\includegraphics[width=5cm]{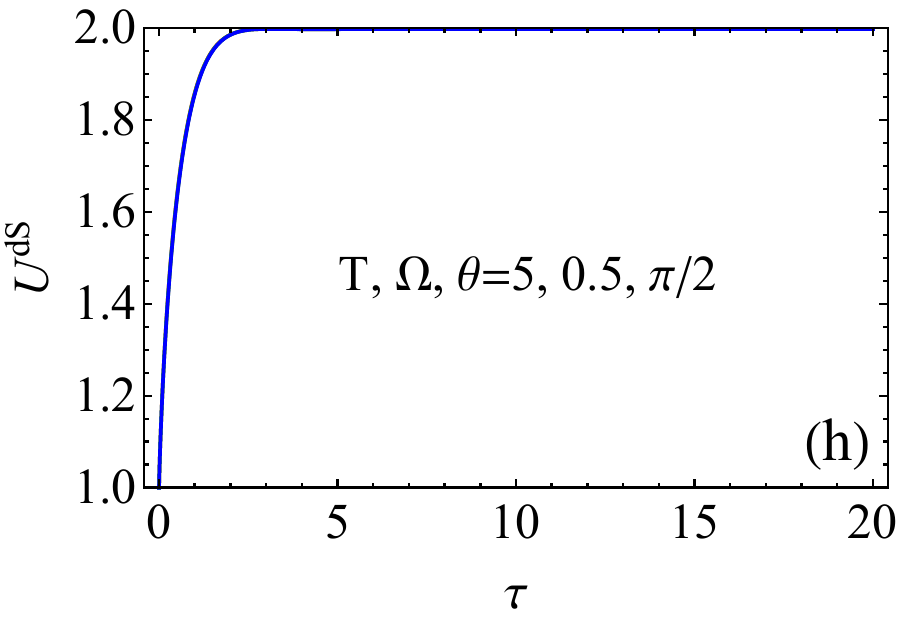}} \ \ \ \
{
\includegraphics[width=5cm]{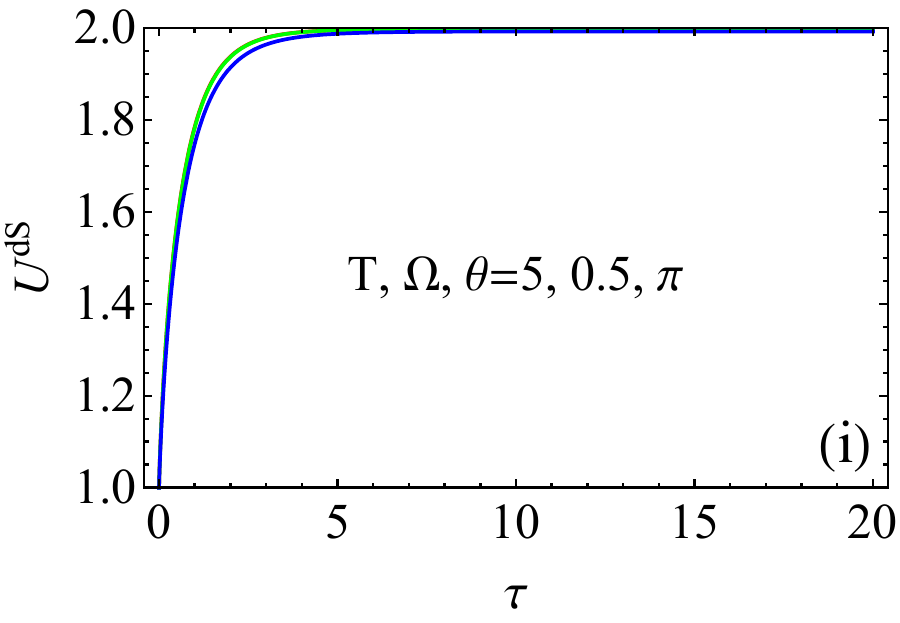}}\ \ \ \
{
\includegraphics[width=5cm]{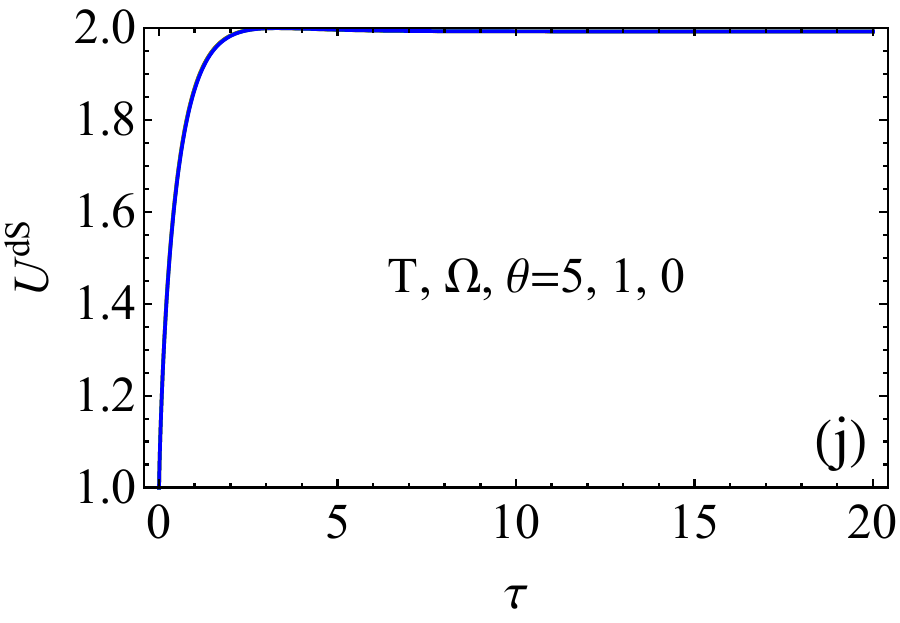}}\ \ \ \
{
\includegraphics[width=5cm]{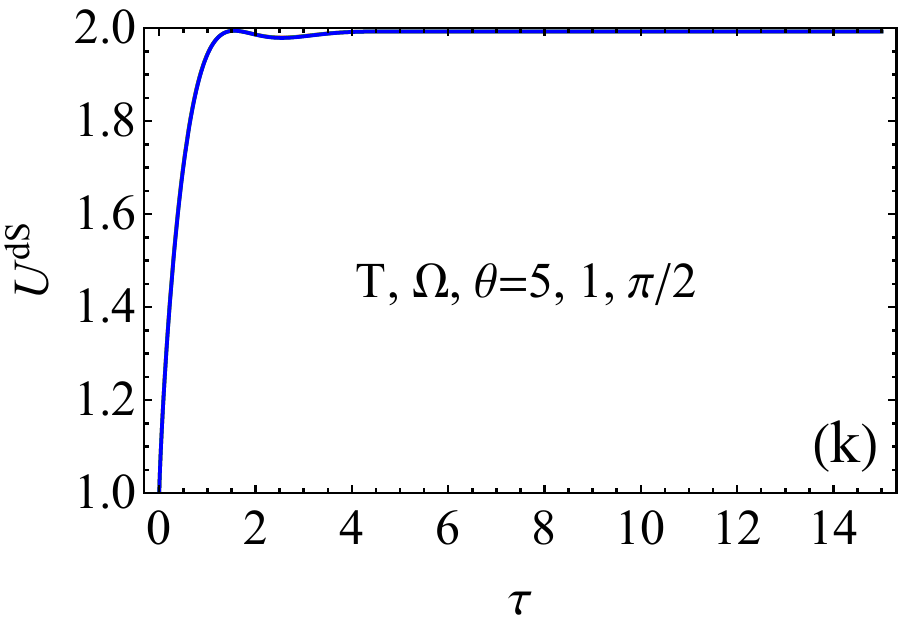}}\ \ \ \
{
\includegraphics[width=5cm]{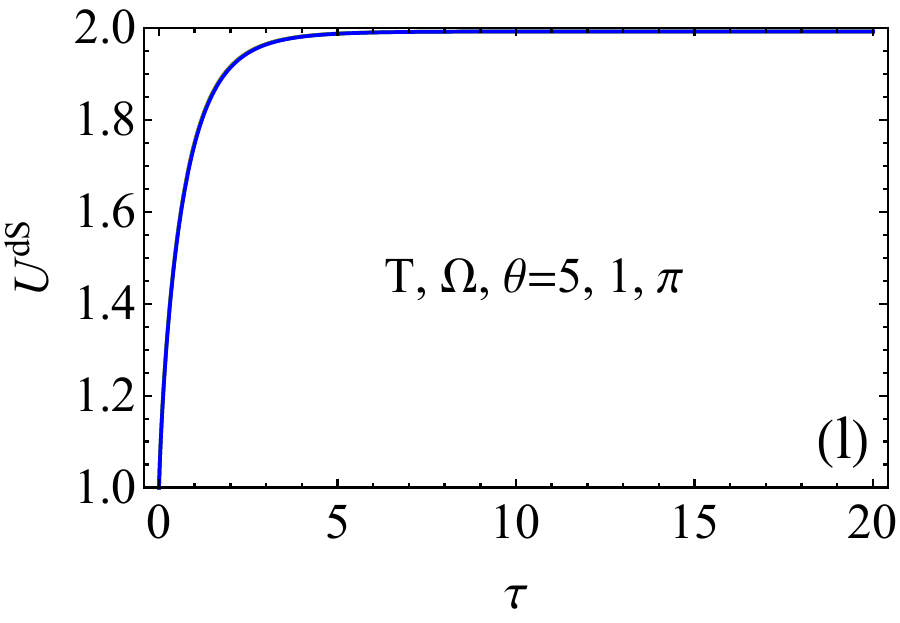}}
\caption{The effect of acceleration on the measurement uncertainty $U^{dS}$ in dS spacetime with different Gibbons-Hawking temperatures $T$, energy gap $\Omega$, and state parameter $\theta$. The different coloured lines represent different acceleration, the red solid line represent $a=0$, the green line represent $a=2$, and the blue line represent $a=3$.}
\label{fig.2}
\end{figure*}
In this section, we will discuss the measured  uncertainty and quantum coherence of the {UDW detector} in dS spacetime. Specifically, we mainly focus on investigating the effects of temperature $T$, energy gap $\Omega$ and acceleration $a$ on dynamics of the uncertainty and coherence of the comoving and accelerated detectors, by resorting to the Pauli operators $\sigma_1$ and $\sigma_3$ as the incompatible measurement.
Incidentally, the initial state $|\psi\rangle$ is highly dependent  of $\theta$.
When $\theta=0$, the initial state is in the ground state $\left| 0 \right\rangle $; when $\theta=\pi$, the initial state becomes in the excited state $\left| 1 \right\rangle $, and the initial state will be in the superposition state of $\frac{1}{{\sqrt 2 }}(\left| 0 \right\rangle  + \left| 1 \right\rangle )$ with   $\theta=\frac{\pi }{2}$.

{Combining Eqs. (\ref{f2}), (\ref{f4}) and (\ref{t8}), one obtains the entropic uncertainty $U^{dS}$ and coherence $C^{dS}$ detected in dS spacetime by comving detector with $a=0$, which have been plotted in Fig. 1 with the growing time $\tau$ for different temperatures $T$ and energy gaps $\Omega$.}
From Fig. \ref{fig.1},  the following conclusions can be drawn: (1) Both entropic uncertainty and coherence will eventually reach { constant} value as time evolves (In order to simplify, {the constant} values of the entropic uncertainty and coherence are denoted as $U_s$ and  $C_s$ respectively, hereafter). (2) The shorter time for uncertainty and coherence approaching the { constant} values $U_s$ and  $C_s$  with the higher $T$. { This is because that the increase of temperature leads to the enhancement of the effect of thermal radiation on the interaction between the detector and the quantum field, which disrupts the original quantum coherence, and the uncertainty of the system increases accordingly. (3) The} increasing $T$ raises the systemic uncertainty and reduces the coherence. Conversely, the increasing $\Omega$ lowers the uncertainty and enhances coherence. (4) Following Figs. \ref{fig.1} (a)-(b) and (c)-(d), it is found that the uncertainty and coherence show an anti-correlated relationship.

\begin{figure*}
\centering
{
\includegraphics[width=5cm]{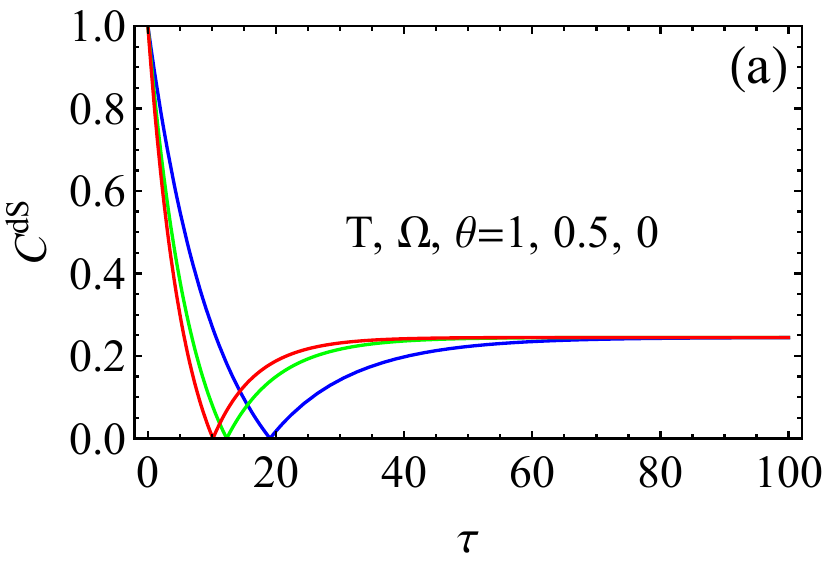}} \ \ \ \
{
\includegraphics[width=5cm]{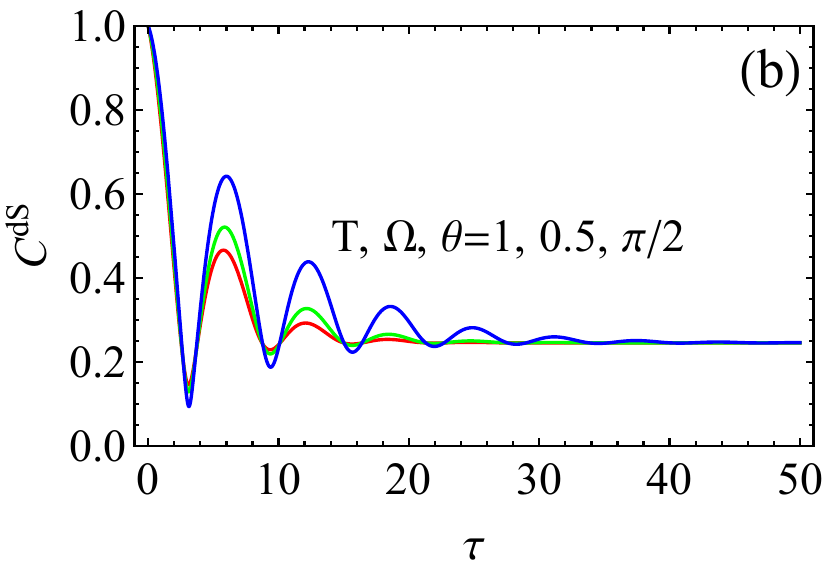}} \ \ \ \
{
\includegraphics[width=5cm]{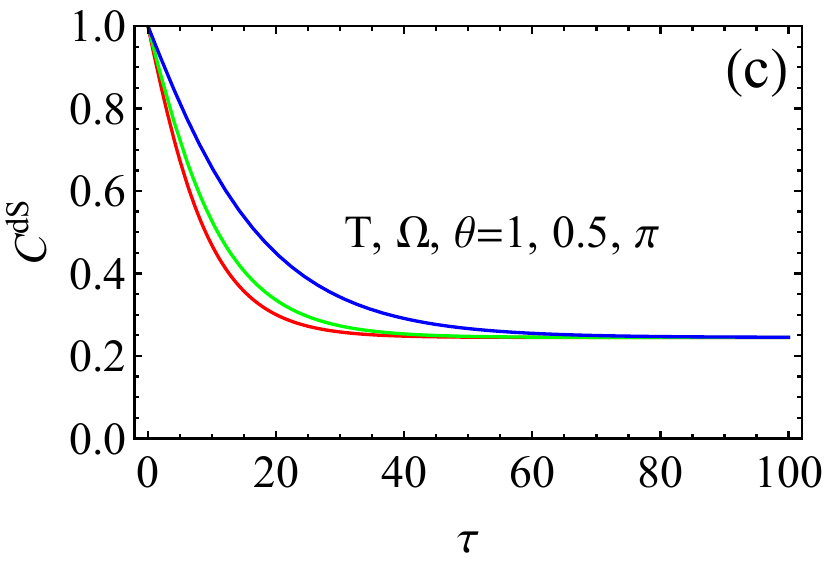}} \ \ \ \
{
\includegraphics[width=5cm]{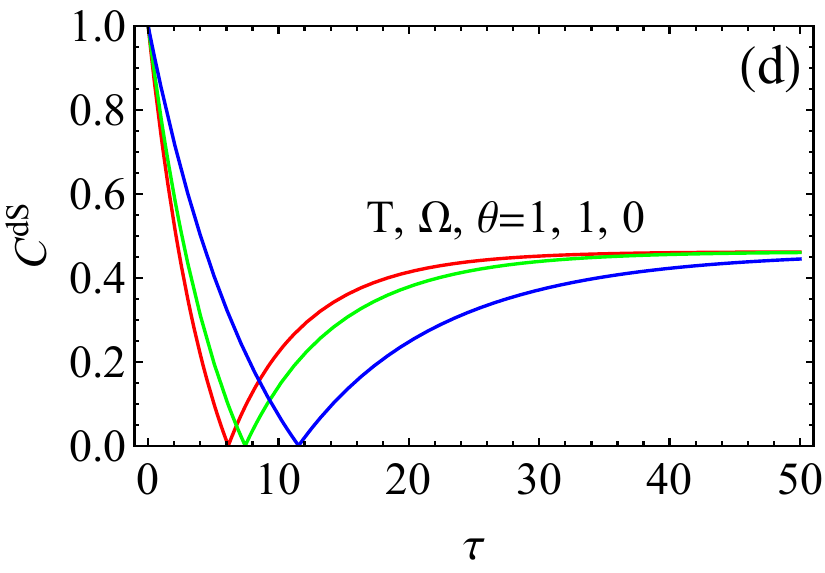}} \ \ \ \
{
\includegraphics[width=5cm]{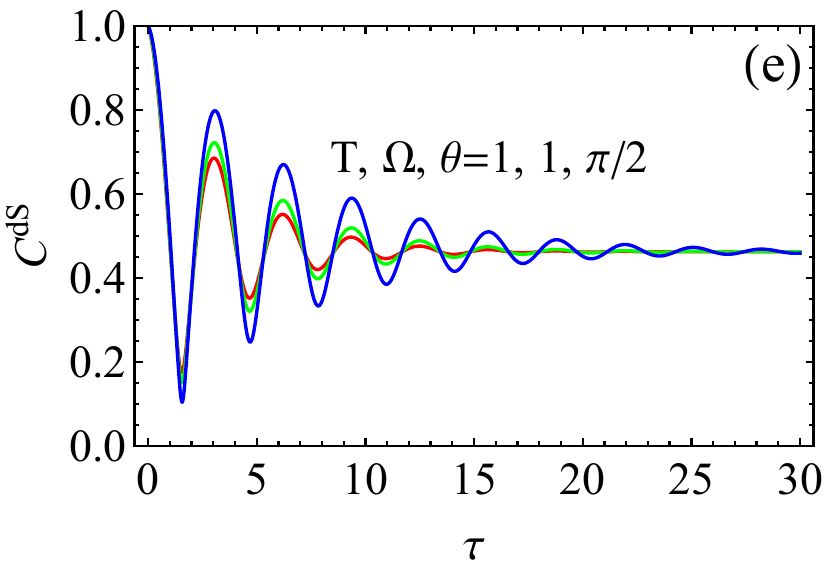}}\ \ \ \
{
\includegraphics[width=5cm]{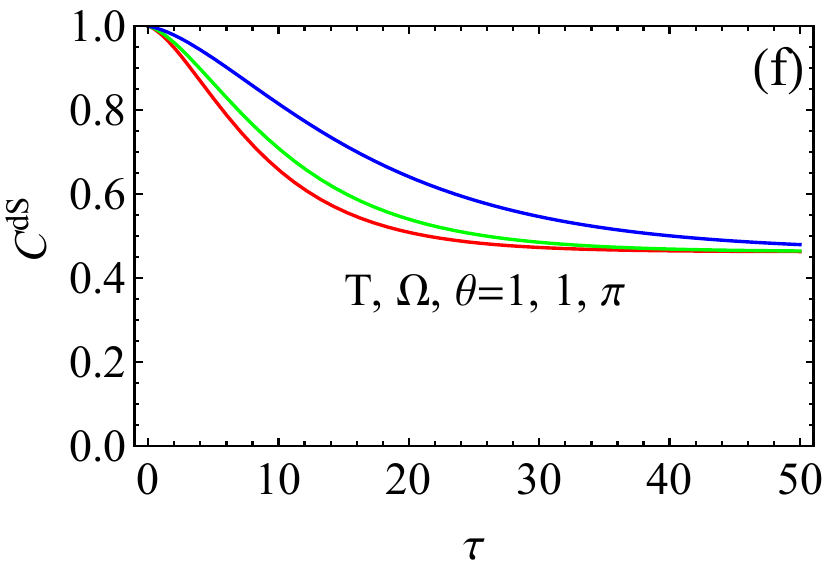}} \ \ \ \
{
\includegraphics[width=5cm]{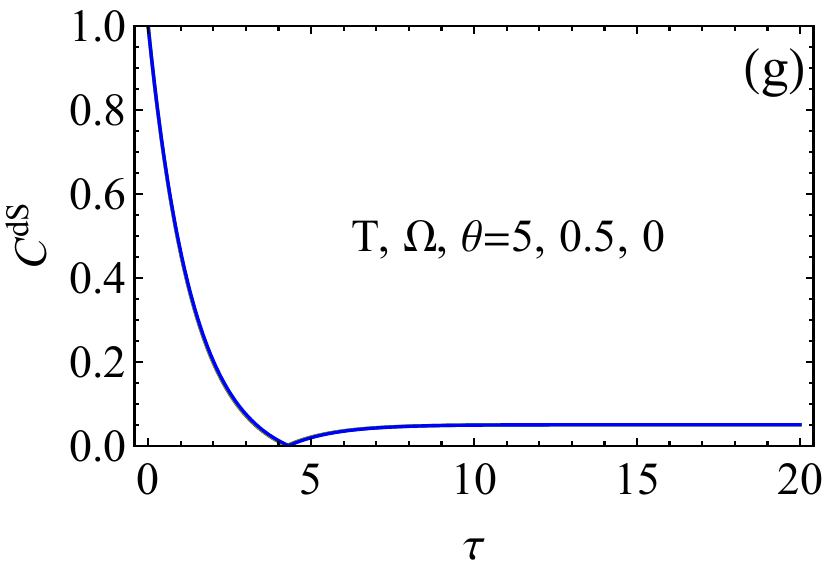}}\ \ \ \
{
\includegraphics[width=5cm]{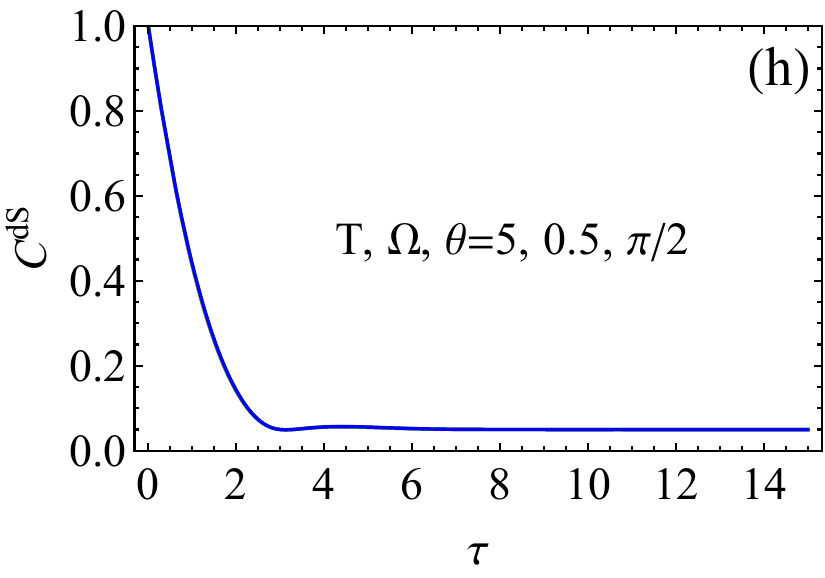}} \ \ \ \
{
\includegraphics[width=5cm]{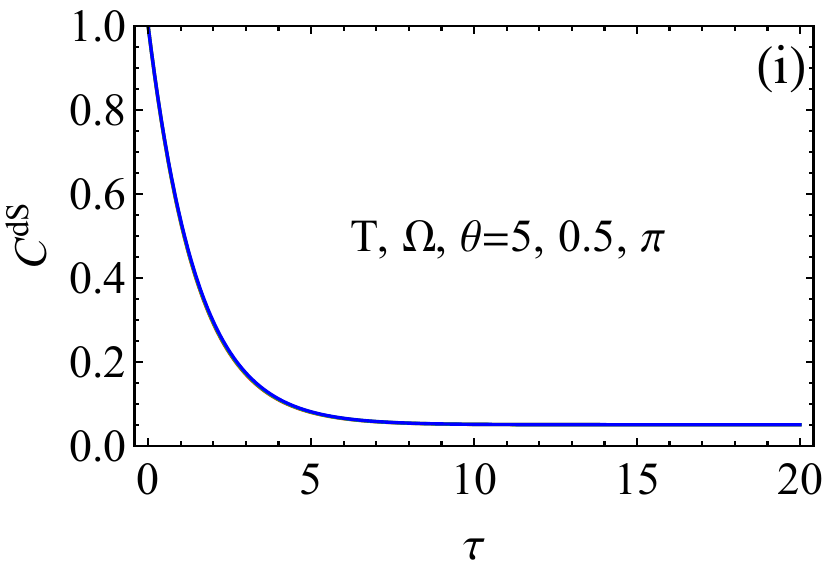}}\ \ \ \
{
\includegraphics[width=5cm]{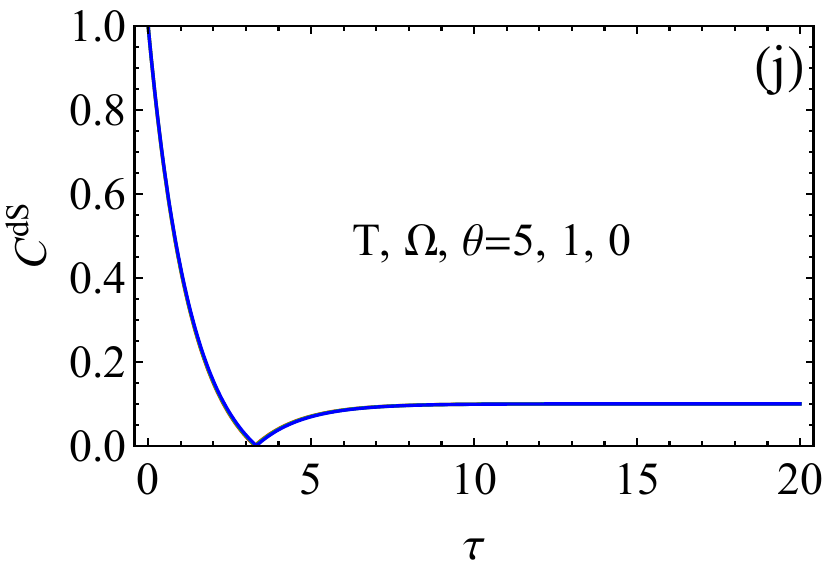}}\ \ \ \
{
\includegraphics[width=5cm]{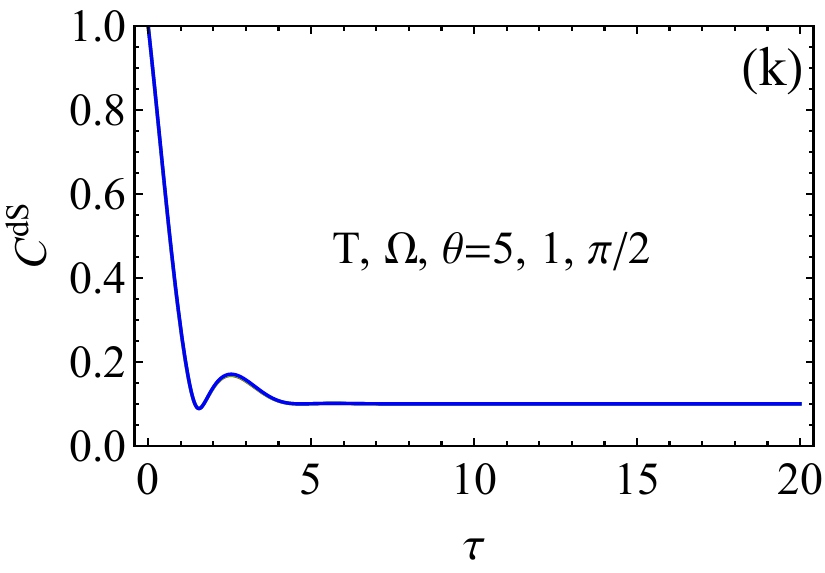}}\ \ \ \
{
\includegraphics[width=5cm]{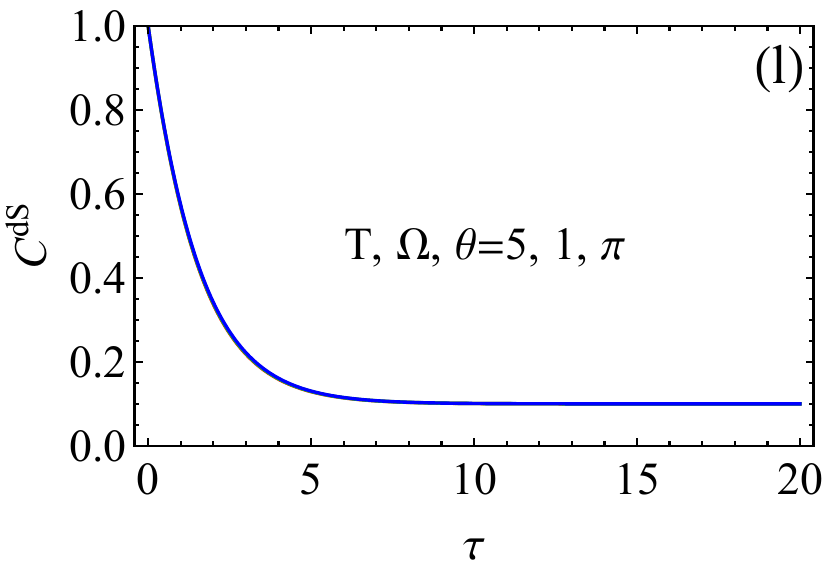}}
\caption{The effect of acceleration on the quantum coherence $C^{dS}$ in dS spacetime with different Gibbons-Hawking temperatures $T$, energy gap $\Omega$, and state parameter $\theta$. The different coloured lines represent different acceleration, the red solid line represent $a=0$, the green line represent $a=2$, and the blue line represent $a=3$.}
\label{fig.3}
\end{figure*}

In fact, the comoving detector discussed above is a detector without acceleration. So, if an acceleration of the detector appears, what effect does the acceleration have on the uncertainty and coherence?
With this in mind, we turn to analyze the case with the accelerated detectors.
{For accelerated detector, the explicit expression of measurement uncertainty $U^{dS}$ can be obtained as}
\begin{align}
U^{dS}&=H^{dS}(\hat{Q})+H^{dS}(\hat{R})  \nonumber\\
&=\frac{{ - 1}}{{2\ln 2}}(\eta_1 \ln \frac{\eta_1 }{2} + \nu_1 \ln \frac{\nu_1 }{2})+ \frac{\ln 2}{\ln 4} \nonumber \\
&-\frac{{ {  \ln\frac{1}{2}( 1\!-\! m_1 )\! +\! \ln(1\! +\!m_1)\! +\! 2m_1\ {\rm artanh}m_1 } }}{{\ln 4}},
\end{align}
with
\begin{align}
\begin{array}{l}
\mu_1  = \frac{{ - \tau \Omega \coth \frac{\Omega }{{2T}}\sqrt { - {a^2} + 4\pi {T^2}} }}{{16\pi T}},\\
m_1 = {e^{\mu_1} }\cos\tilde \Omega \tau \sin\theta, \\
\eta_1  = 1 - {e^{\mu_1} }\cos\theta  + (1 - {e^{\mu_1 /2}})\tanh\frac{\Omega }{{2T}},\\
\nu_1  = 1 + {e^{\mu_1} }\cos\theta  - (1 - {e^{\mu_1 /2}})\tanh\frac{\Omega }{{2T}}.
\end{array}
\label{t13}
\end{align}
Relying on Eq. (\ref{f4}), the quantum coherence $C^{dS}$ of the detector's system is expressed as
\begin{align}
C^{dS}= \frac{1}{2}\left| {m_1 + i(\eta_1  - 1)} \right| + \frac{1}{2}\left| {m_1 + i(\nu_1  - 1)} \right|.
\end{align}
For clarity, the measurement uncertainty {and quantum coherence} in de Sitter spacetime with temperature $T$, energy gap $\Omega$, and state parameter $\theta$ {are drawn in Fig. \ref{fig.2} and Fig. \ref{fig.3} }for the accelerated detectors.

{Following  Figs. \ref{fig.1}-\ref{fig.3}, two interesting conclusions are obtained as: (1) The oscillations of the uncertainty magnitude and coherence only take place when the initial state is in a superposition state. (2) The high temperature suppresses the  oscillations. Here, we proceed by revealing the reasons behind these results. Explicitly,
the value of $m_1$ in Eq. (\ref{t13}) is non-zero,  when the initial state is a superposition with $\theta\neq 0\ {\rm or}\ \pi$. Since $\cos\tilde \Omega \tau$ in $m_1$ is a periodic function whose value oscillates between -1 and 1, causing  $m_1$ to oscillate as the  increasing $\tau$, which leads to oscillations of $U^{dS}$ and $C^{dS}$ during the beginning evolution.
This also explains why the oscillatory behavior becomes more pronounced when the energy gap increases. Notably,  $e^{\mu_1}$ is responsible for the system's evolution as $\tau$ increases. Since ${\mu _1} \propto  - \tau$ and ${m_1} \propto {e^{{\mu _1}}}$, we have ${m_1} \to 0$ for $\tau  \to \infty $, which can be used for perfectly explaining the disappearance of oscillating behaviors in the late stages.
As to the second conclusion, this is due to that: $\coth \frac{\Omega }{{2T}} \to \infty $ for large enough $T$, leading to ${\mu _1}\to -\infty $. Since ${m_1} \propto {e^{{\mu _1}}}$, it follows that ${m_1} \to 0$ with increasing $T$. As a result, the oscillatory term $\cos\tilde \Omega \tau$ in ${m_1}$ no longer contributes to the  uncertainty and quantum coherence, causing ${U}^{ds}$ and ${C}^{ds}$ to no longer exhibit oscillatory behavior.
Physically, the thermal noise and thermal excitation of the system increase at high temperatures, the system tends to be in thermal equilibrium, with the oscillations being suppressed.
Thus, it is observed that the oscillation weakens with increasing temperature as shown in Figs.  \ref{fig.2} and \ref{fig.3}.}

{Moreover, the changing acceleration can affect quantum uncertainty and quantum coherence, however, its effect is limited and does not alter the final evolution of the uncertainty and coherence. The effect of $a$ will be negligible for higher $T$. This is because the environmental thermal noise becomes more pronounced for high $T$, and the effects of detector's acceleration are overshadowed by the disturbances from thermal radiation, resulting in disappearance of the effects of detector's acceleration. In addition,
it deserves noting that the measurement uncertainty is anti-correlated with quantum coherence during the evolution.}

{ Furthermore, it is evident  that the detector ultimately thermalizes with the environment as the interaction time increases, leading to that both uncertainty and quantum coherence  approach the constant values ${U}_s^{ds}$ and ${C}_s^{ds}$, as displayed in Figs. \ref{fig.1}, \ref{fig.2} and \ref{fig.3}. Additionally, it can be observed that the higher $T$ results in  larger constant value ${C}_s^{ds}$ and smaller ${U}_s^{ds}$. This naturally raises a question: what factors determine the constant value?} Keeping this in consideration, we investigate the evolution of uncertainty as time approaches infinity in the following. When $\tau  \to \infty $, the final state of the system can be expressed as
\begin{align}
 \rho(\infty)=\mathop {{\lim}}\limits_{\tau  \to \infty } \rho(\tau)  = \left( {\begin{array}{*{20}{c}}
{\frac{1}{{1 + {{\rm{e}}^{ - \frac{\Omega }{T}}}}}}&{ - \frac{1}{2}{i}\tanh (\frac{\Omega }{{2T}})}\\
{\frac{1}{2}{i}\tanh (\frac{\Omega }{{2T}})}&{\frac{1}{{1 + {{\rm{e}}^{\frac{\Omega }{T}}}}}}
\end{array}} \right).
\end{align}
Therefore, {the constant value }${U}_s^{dS}$ of systemic state $ \rho(\infty)$ can be expressed as
\begin{align}
{{U}_s^{dS}(\tau  \to\infty)=\frac{{{{\rm{e}}^{\frac{\Omega }{T}}}( {\ln  2 - \ln \frac{1}{{1 + {{\rm{e}}^{ - \frac{\Omega }{T}}}}}} ) - \ln \frac{1}{{2 + 2{{\rm{e}}^{\frac{\Omega }{T}}}}}}}{{(1 + {{\rm{e}}^{\frac{\Omega }{T}}})\ln 2}}.}
\label{t11}
\end{align}
Consequently, it can be   inferred from Eq. (\ref{t11}) that the uncertainty ${U}_s^{dS}$  is independent of the acceleration $a$ and the state's parameter $\theta$. More specifically, it is only related to the ratio of the temperature $T$ and energy gap $\Omega$, namely, $\Omega/T$.

\begin{figure*}
\centering
{
\includegraphics[width=6.5cm]{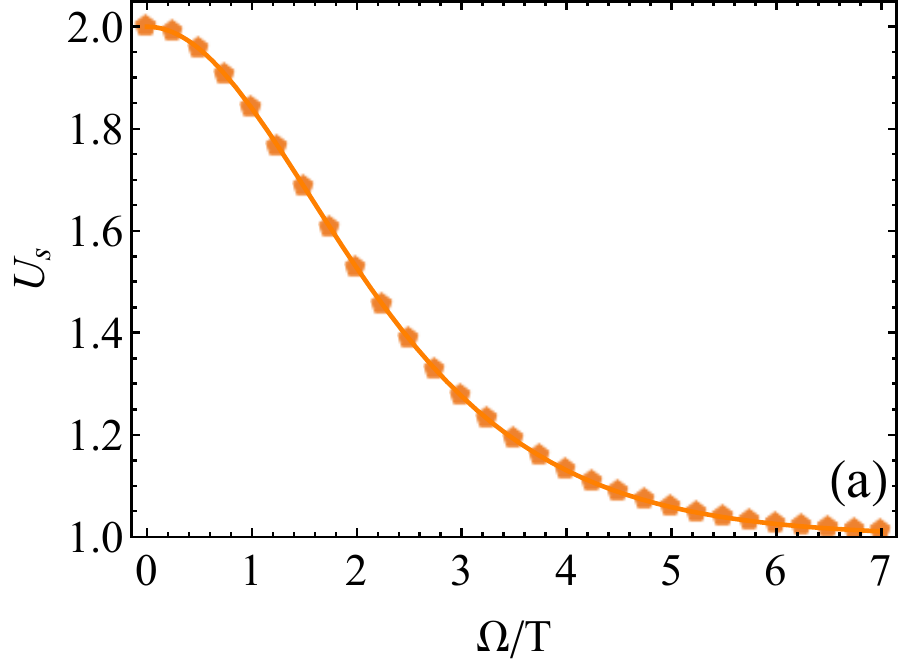}} \ \ \ \
{
\includegraphics[width=6.5cm]{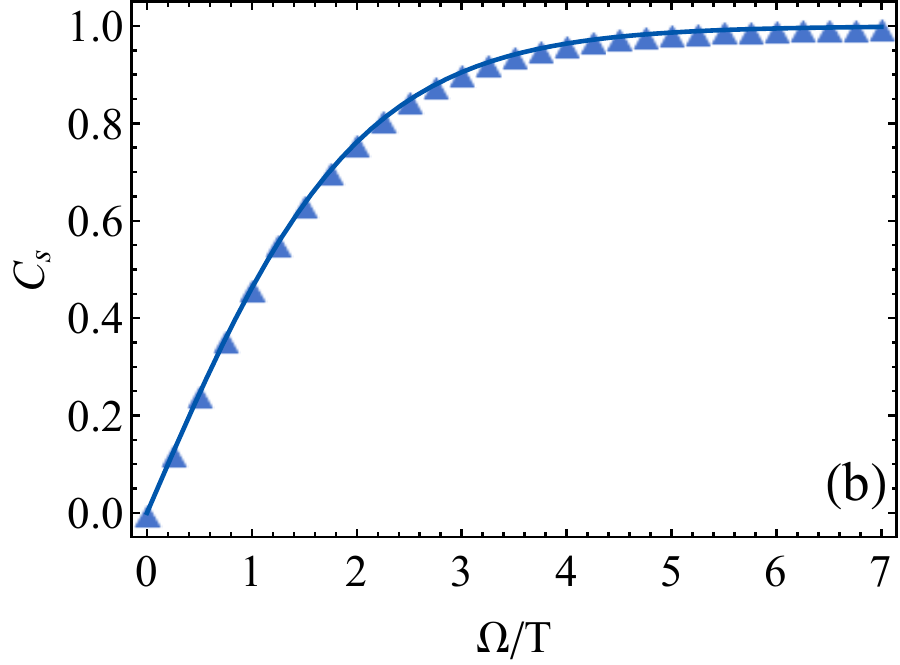}}
\caption{(a) The { constant} value of measurement uncertainty $U_s$ vs the ratio $\Omega/T$. (b) The { constant} value of coherence $C_s$ vs the ratio $\Omega/T$.}
\label{fig.4}
\end{figure*}

Likewise, the system's coherence also tends to a { constant} value when $\tau  \to \infty $. Ultimately, the expression of $C_s^{dS}$  can be formulated as
\begin{align}
{{C}_s^{ds}(\tau  \to\infty) = \left| { - \frac{1}{2}{i}\tanh (\frac{\Omega }{{2T}})} \right| + \left| {\frac{1}{2}{i}\tanh (\frac{\Omega }{{2T}})}\right|.}
\label{t12}
\end{align}
It is evident that ${C}_s^{dS}$   also relies on the ratio $\Omega/T$ of the temperature and energy gap from Eq. (\ref{t12}).
More detailedly, the evolution of $U_s$ and $C_s$ with the ratio $\Omega/T$ has been plotted as Fig. \ref{fig.4}. Following the figure, it is evident that  ${U}_s^{dS}$ gradually  decreases progressively to 1 as the ratio increases, while ${C}_s^{dS}$ steadily approaches its maximum. At the same time, it is interesting to see  that ${U}_s^{dS}$ and  ${C}_s^{dS}$ exhibit an inverse correlation, as shown in  Figs. \ref{fig.4} (a) and (b).
This can be interpreted as that the stronger quantumness will induce less measurement uncertainty in the current architecture, and vice versa.

\section{Measurement Uncertainty and Quantum coherence in Anti-de Sitter Spacetime}

\begin{figure*}
\centering
{
\includegraphics[width=5cm]{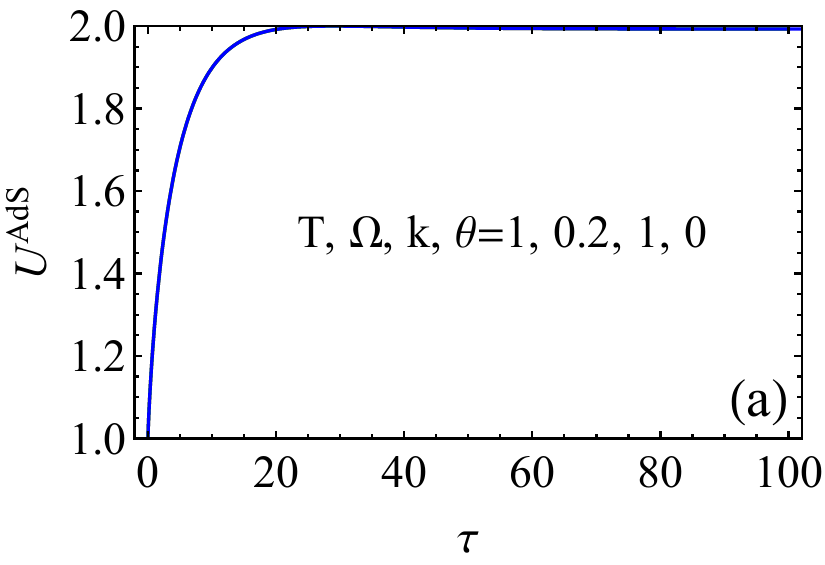}}\ \ \ \
{
\includegraphics[width=5cm]{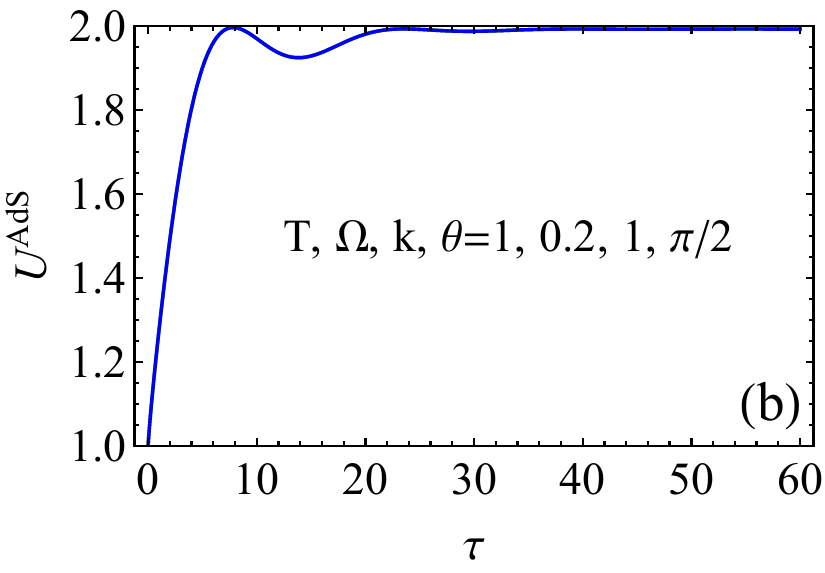}}
{
\includegraphics[width=5cm]{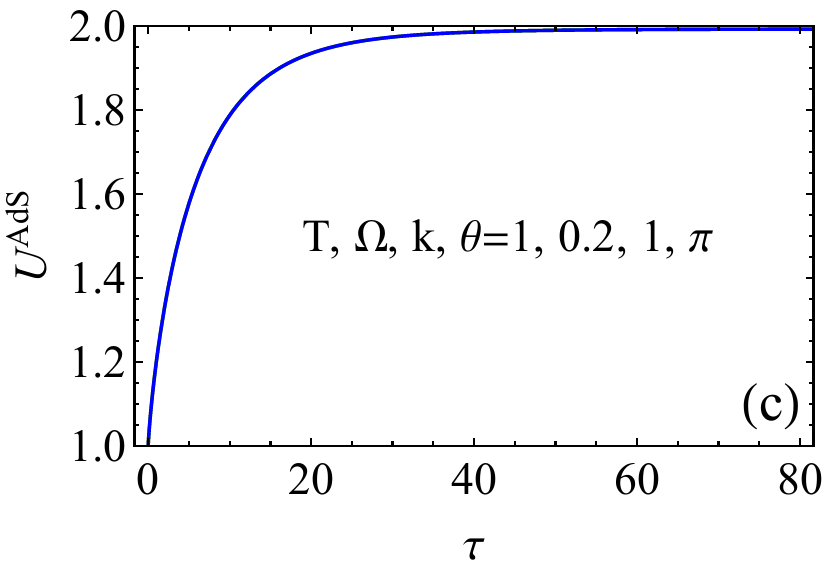}} \ \ \ \
{
\includegraphics[width=5cm]{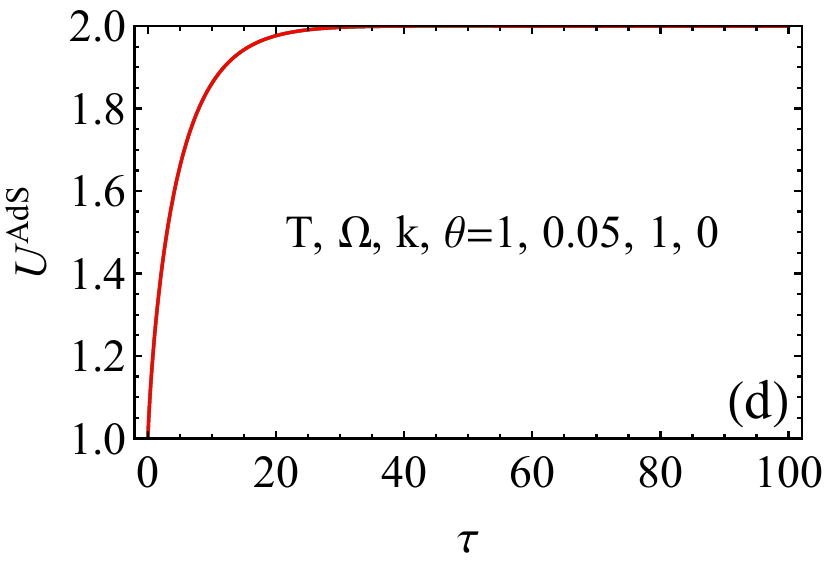}} \ \ \ \
{
\includegraphics[width=5cm]{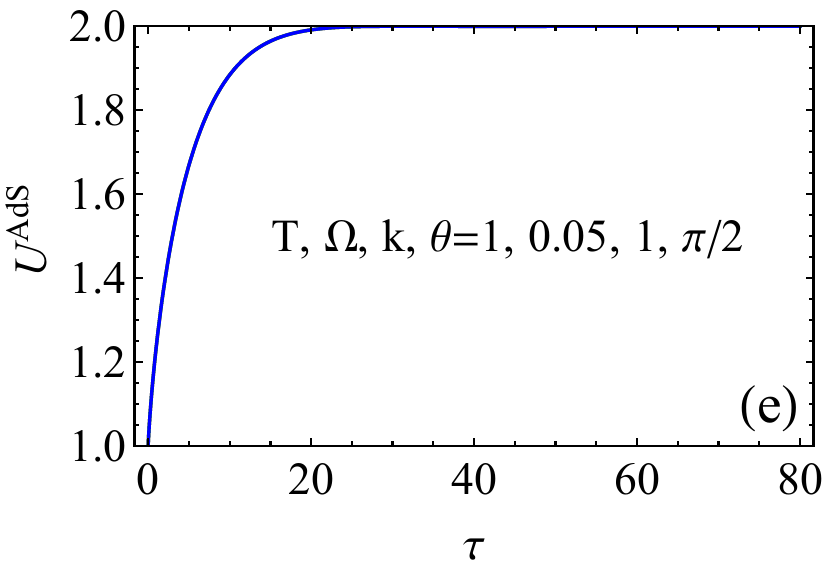}} \ \ \ \
{
\includegraphics[width=5cm]{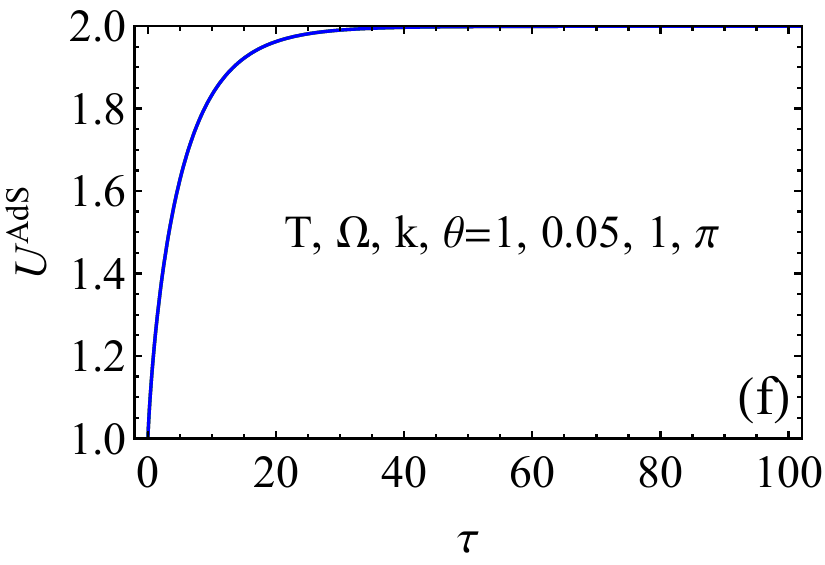}}\ \ \ \
{
\includegraphics[width=5cm]{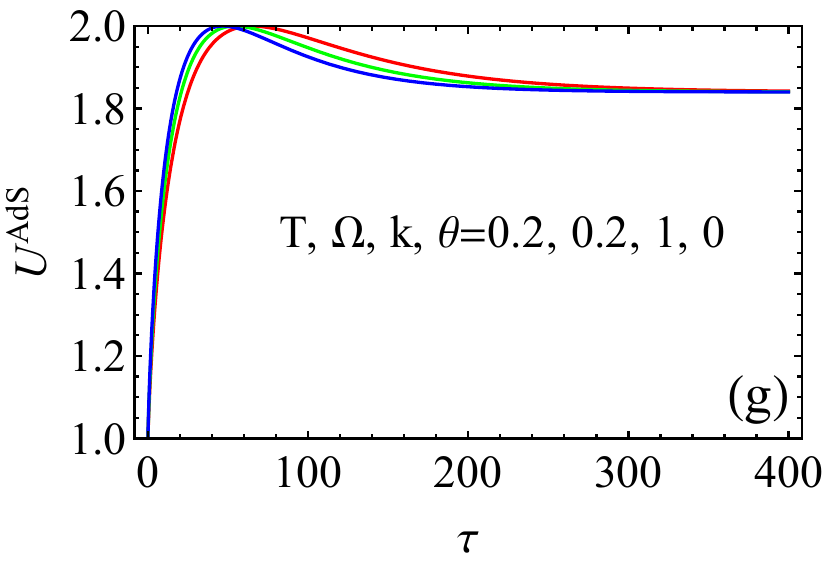}} \ \ \ \
{
\includegraphics[width=5cm]{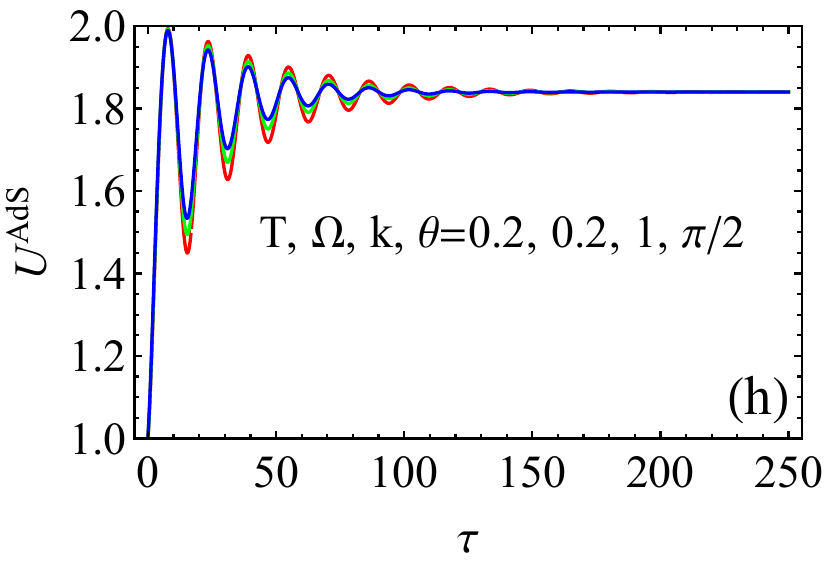}}\ \ \ \
{
\includegraphics[width=5cm]{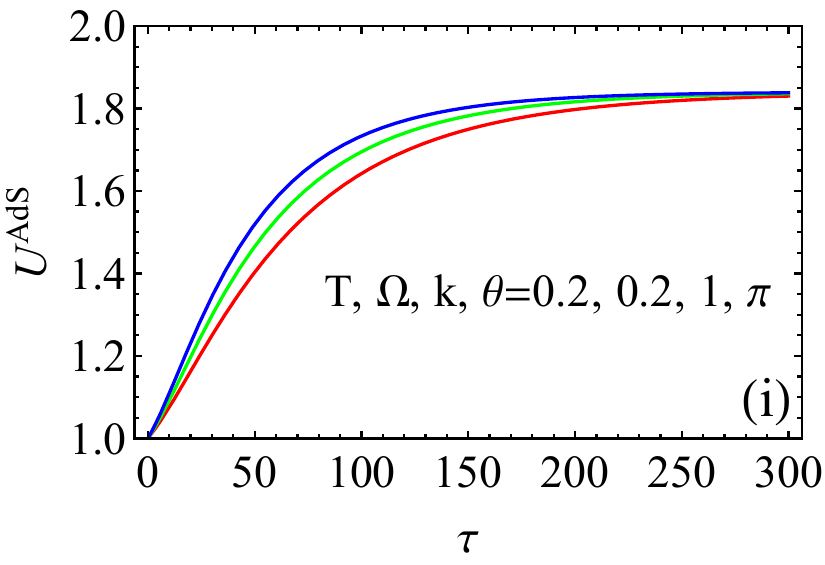}} \ \ \ \
{
\includegraphics[width=5cm]{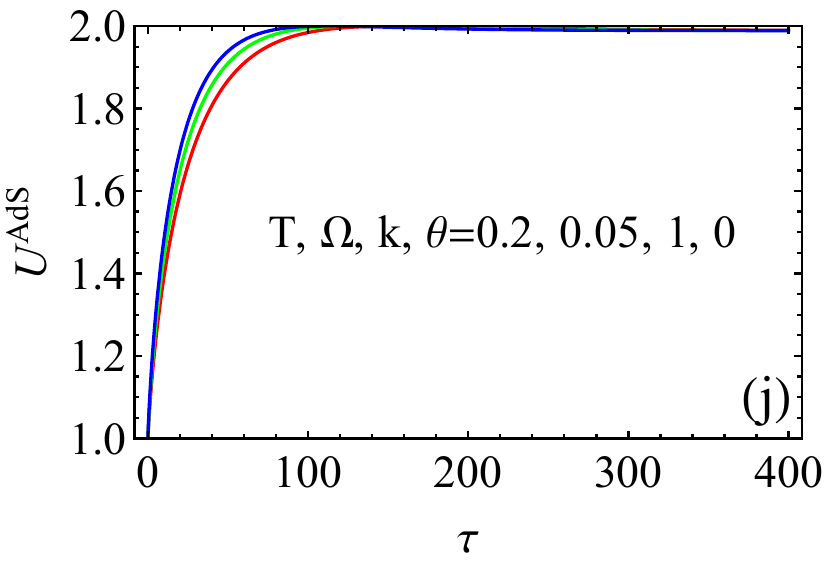}}\ \ \ \
{
\includegraphics[width=5cm]{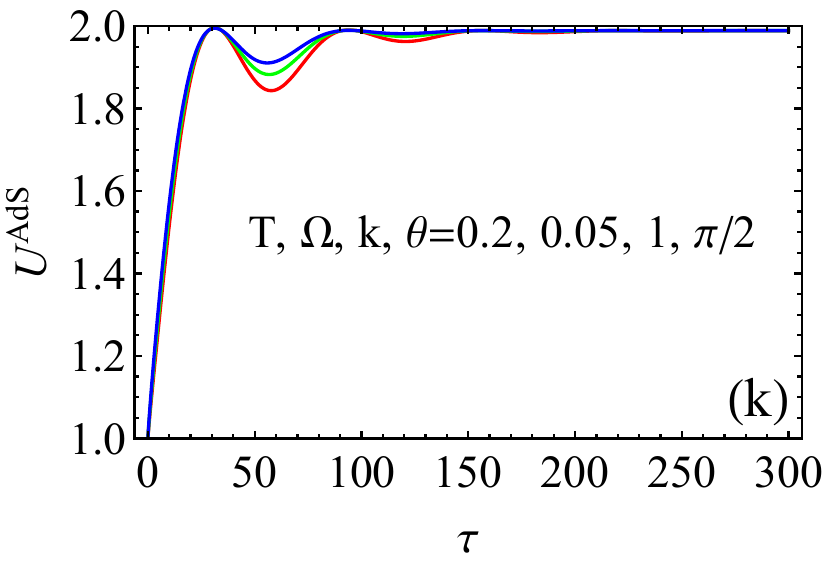}} \ \ \ \
{
\includegraphics[width=5cm]{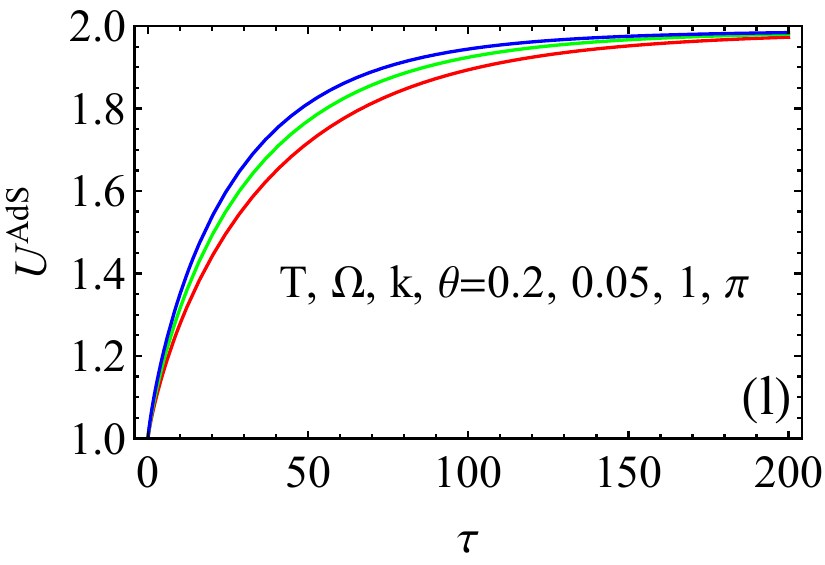}}
\caption{The effect of  boundary condition on the measurement uncertainty $U^{AdS}$ in AdS spacetime with different temperatures $T$, energy gap $\Omega$, the curvature $k$, and state parameter $\theta$. The different coloured lines represent different boundary conditions, the red solid line represent Dirichlet boundary condition ($\zeta= 1$), the green line represent transparent boundary condition ($\zeta= 0$), and the blue line represent Neumann boundary condition ($\zeta=-1$). Some of the pictures have only one line because three different coloured lines overlap.}
\label{fig.5}
\end{figure*}
Next, let us turn to explore the dynamics of the measurement uncertainty $U^{AdS}$ and quantum coherence $C^{AdS}$ in AdS spacetime in this section, focusing on the effects of three different boundary conditions, temperature $T$, initial state $\theta$, and energy gap $\Omega$ on $U^{AdS}$ and $C^{AdS}$.
The effect of the acceleration $a$ of the detector is also taken into account. For simplicity, we use the curvature $k$ and temperature $T$ instead of $a$, since it is necessary to ensure that $k = \sqrt {{a^2} - 4{\pi ^2}{T^2}} $ for real in AdS spacetime.

Similarly, we obtain the   expression for  the uncertainty of interest in AdS spacetime by Eq. (\ref{f2}), namely,
\begin{align}
U^{AdS}&=H^{AdS}(\hat{Q})+H^{AdS}(\hat{R})\nonumber\\
&=1\! +\frac{{ - 1}}{{2\ln 2}}({\eta _2}\ln \frac{{{\eta _2}}}{2} + {\nu _2}\ln \frac{{{\nu _2}}}{2}) \nonumber\\
&-\! \frac{ {\ln(1\! -\! {m_2})\! +\! \ln(1 \!+\! {m_2}) \!+\! 2{m_2}{{{\rm arctanh} }m_2}} }{{\ln 4}},
\end{align}
with
\begin{align}
\begin{array}{l}
{\mu _2} = \frac{{ - \tau \coth \frac{\Omega }{{2T}}[ - 2\Omega  + \frac{{k\zeta \sin (\frac{{\Omega {\rm csch}\frac{{2\pi T}}{k}}}{{\pi T}})}}{{\sqrt {1 + \frac{{4{\pi ^2}T}}{k}} }}]}}{{16\pi T}},\\
{m_2} = {e^{{\mu _2}}}\cos\tilde \Omega \tau \sin\theta ,\\
{\eta _2} = 1 - {e^{{\mu _2}}}\cos\theta  + (1 - {e^{{\mu _2}/2}})\tanh\frac{\Omega }{{2T}},\\
{\nu _2} = 1 + {e^{\mu _2} }\cos\theta  - (1 - {e^{{\mu _2} /2}})\tanh\frac{\Omega }{{2T}}.
\end{array}
\end{align}
And the quantum coherence can be given as
\begin{align}
C^{AdS}= \frac{1}{2}\left| {m_2+ i(\nu _2  - 1)} \right| + \frac{1}{2}\left| {m_2 + i(\eta_2  - 1)} \right|
\end{align}
in the current scenario.

\begin{figure*}
\centering
{
\includegraphics[width=5cm]{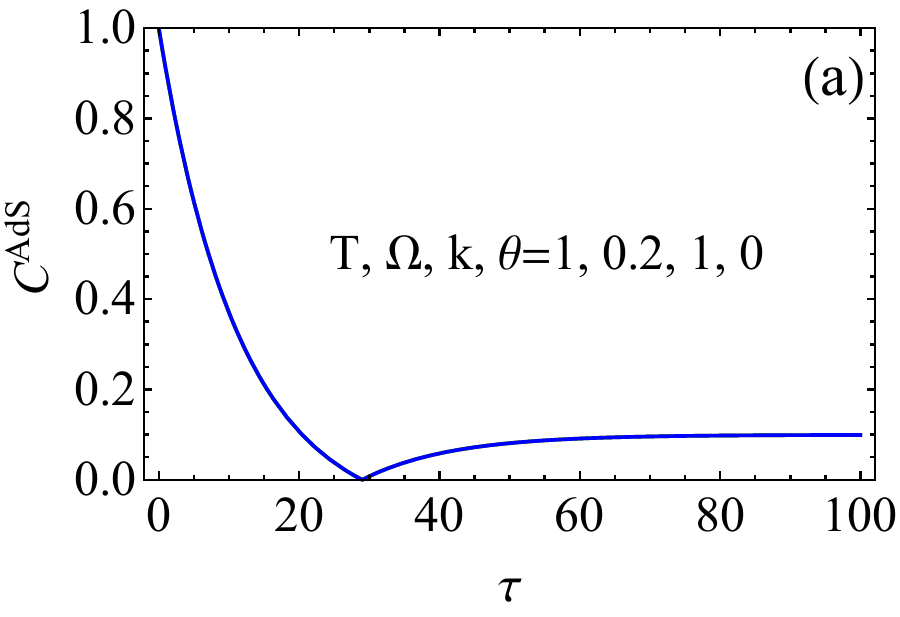}}\ \ \ \
{
\includegraphics[width=5cm]{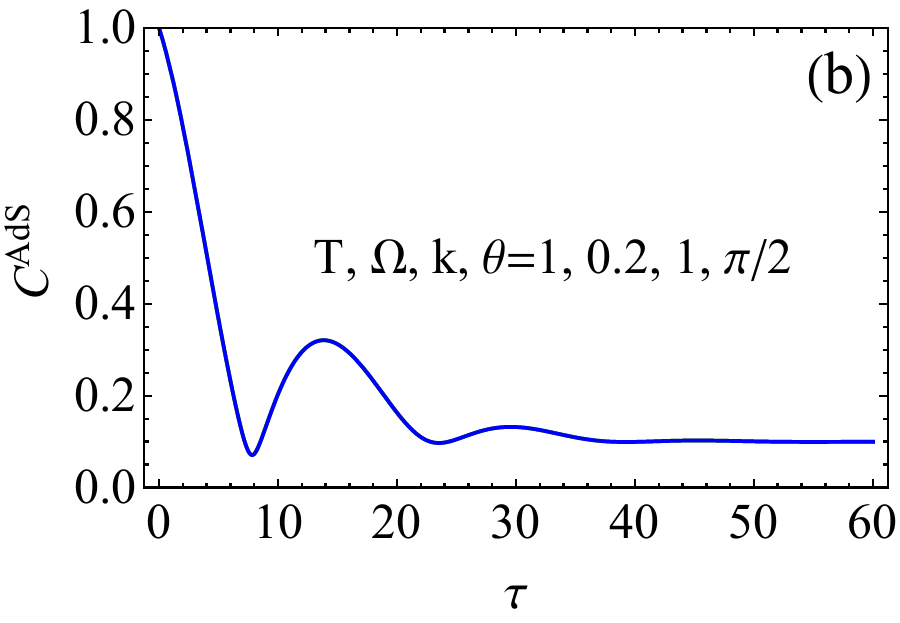}}
{
\includegraphics[width=5cm]{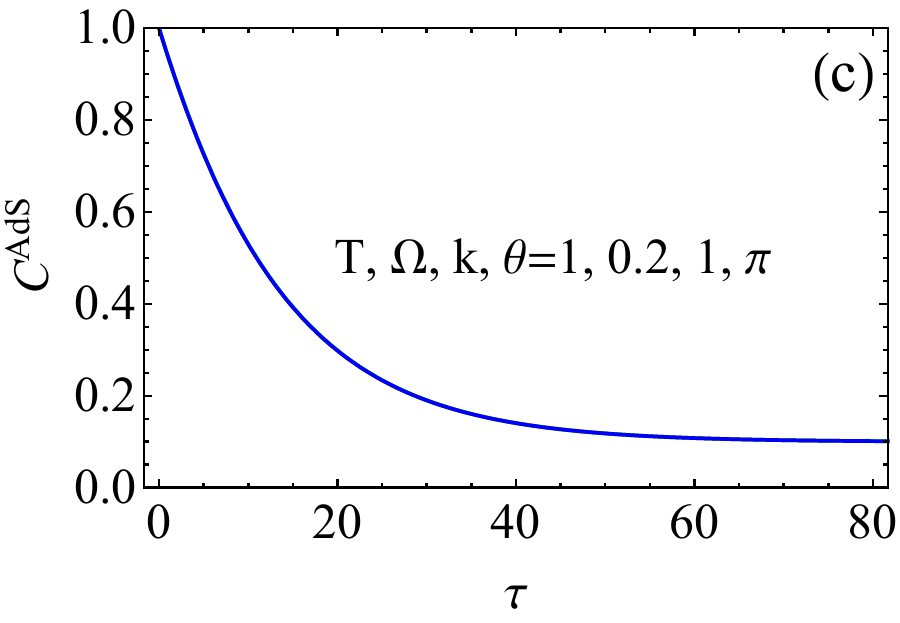}} \ \ \ \
{
\includegraphics[width=5cm]{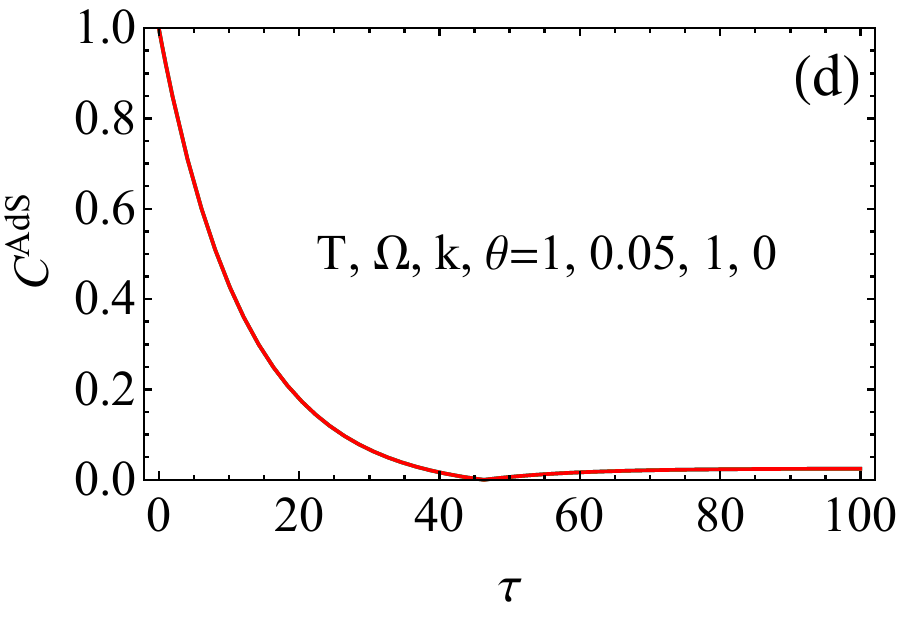}} \ \ \ \
{
\includegraphics[width=5cm]{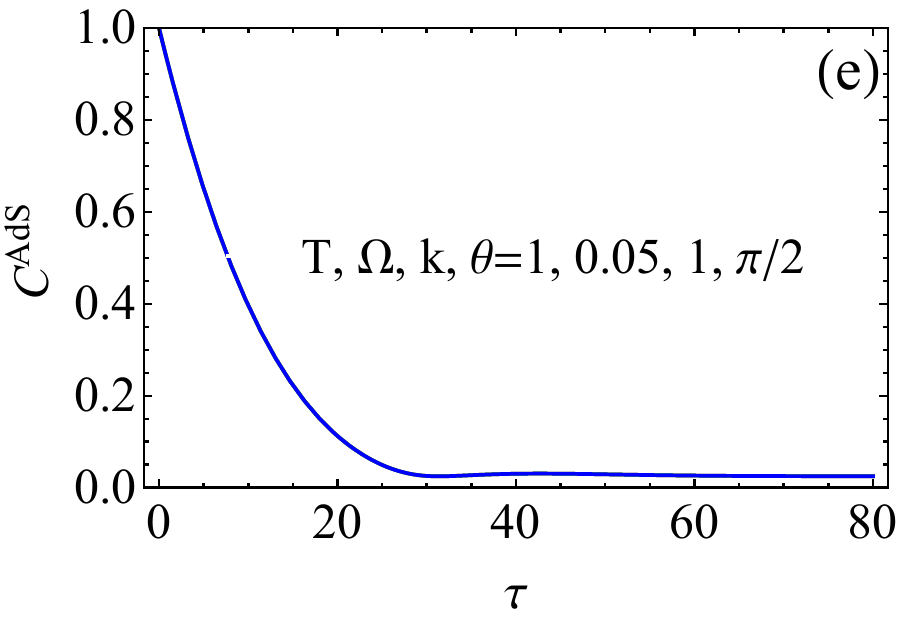}} \ \ \ \
{
\includegraphics[width=5cm]{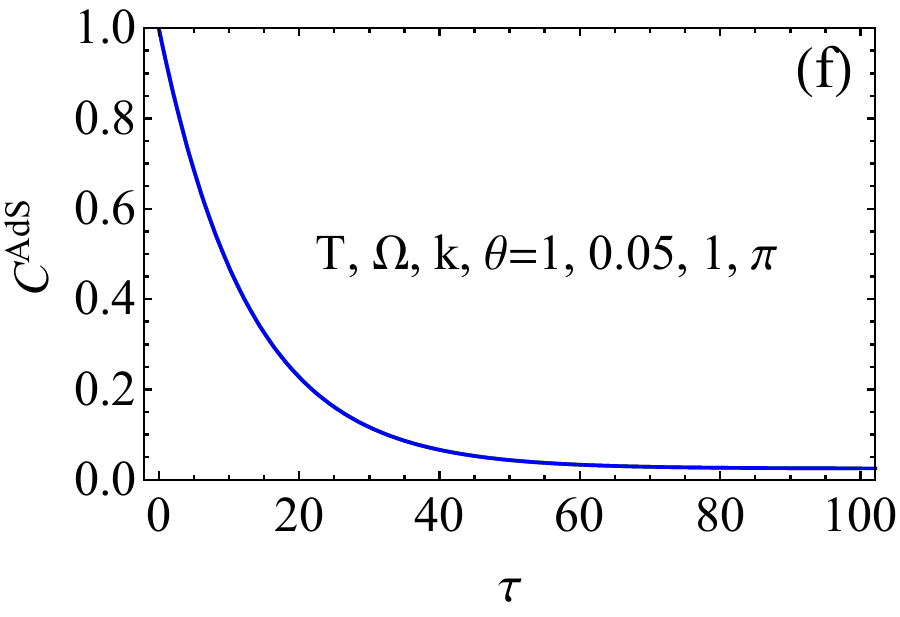}}\ \ \ \
{
\includegraphics[width=5cm]{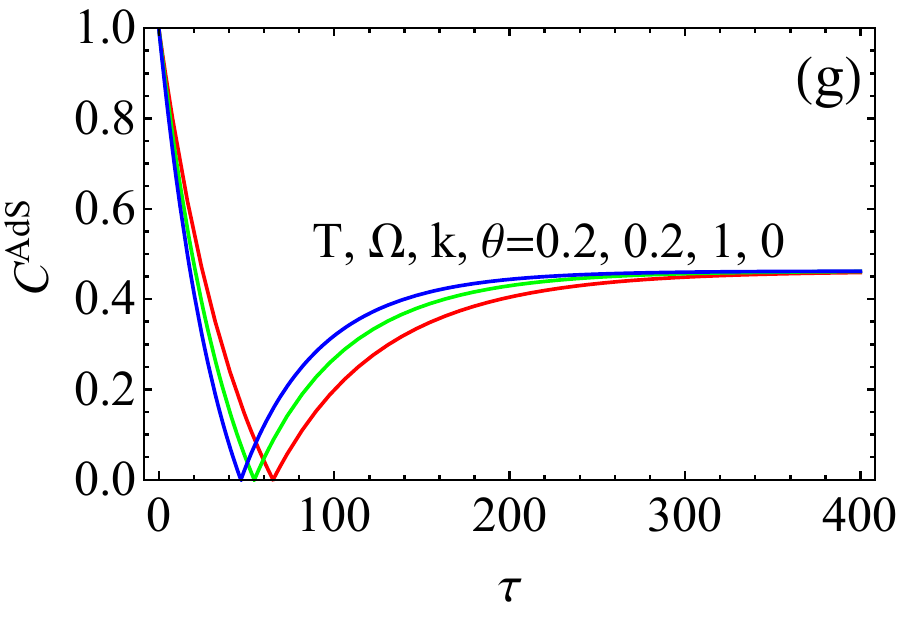}} \ \ \ \
{
\includegraphics[width=5cm]{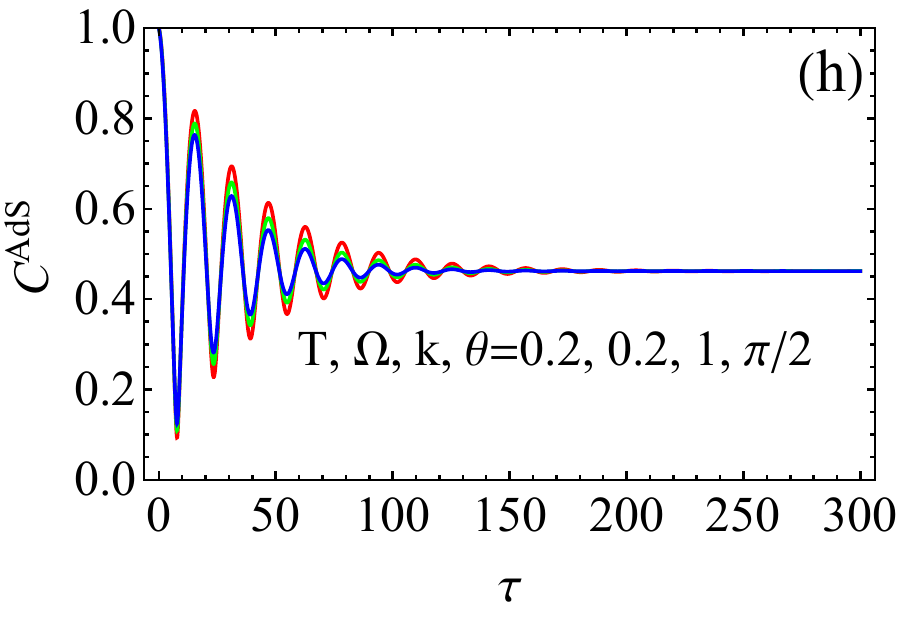}}\ \ \ \
{
\includegraphics[width=5cm]{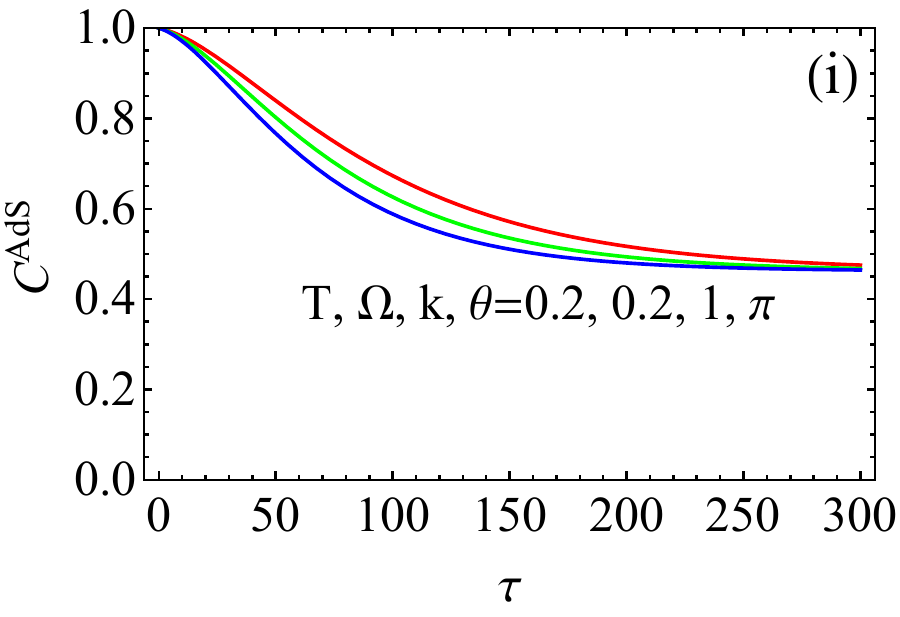}} \ \ \ \
{
\includegraphics[width=5cm]{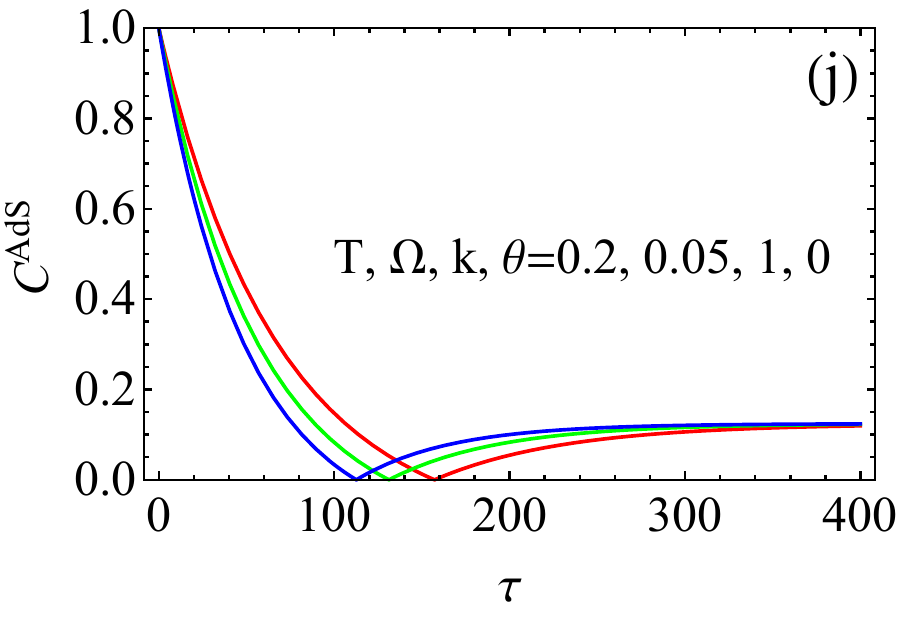}}\ \ \ \
{
\includegraphics[width=5cm]{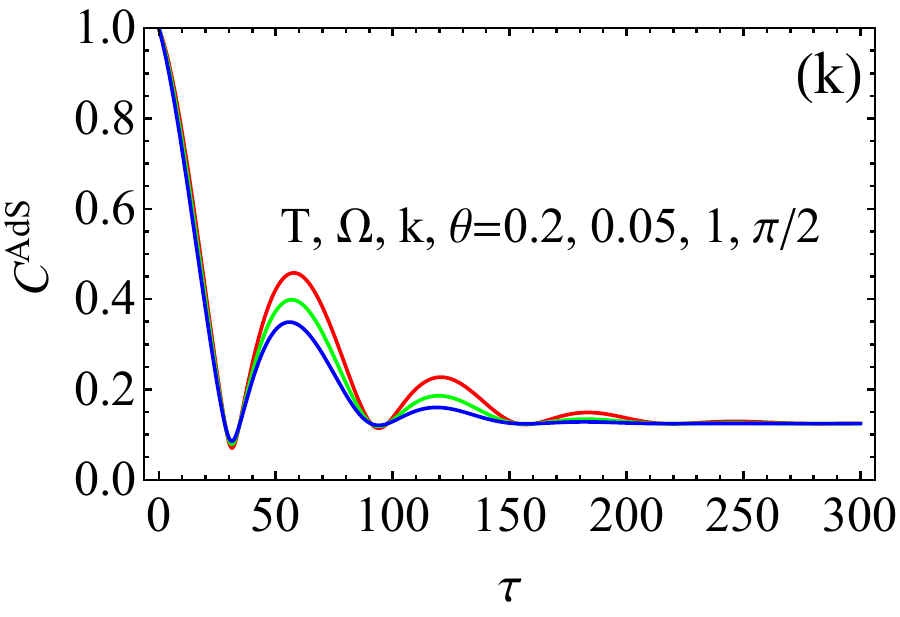}} \ \ \ \
{
\includegraphics[width=5cm]{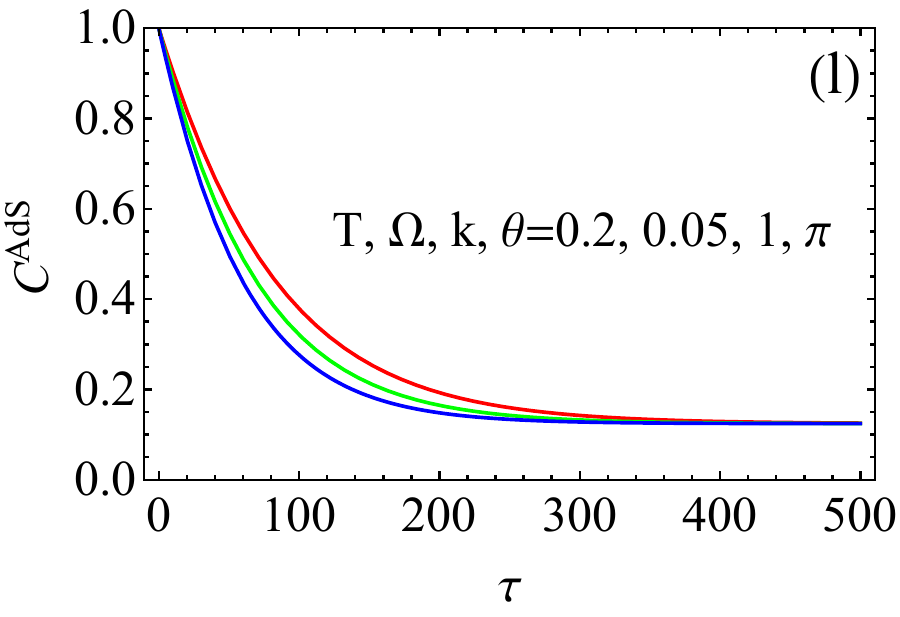}}
\caption{The effect of  boundary condition on the quantum coherence $C^{AdS}$ in AdS with different temperatures $T$, energy gap $\Omega$, the curvature $k$, and state parameter $\theta$. The different coloured lines represent different boundary conditions, the red solid line represent Dirichlet boundary condition ($\zeta= 1$), the green line represent transparent boundary condition ($\zeta= 0$), and the blue line represent Neumann boundary condition ($\zeta=-1$). Some of the pictures have only one line because three different coloured lines overlap.}
\label{fig.6}
\end{figure*}

The time evolution of uncertainty {and quantum coherence }under the three boundary conditions $\zeta$ with different $T$, $k$,  $\theta$ and $\Omega$ {have been illustrated in Figs. \ref{fig.5} and  \ref{fig.6}, respectively.}
The uncertainty and quantum coherence at different boundary conditions eventually reach a { constant} value under the time evolution.  This is attributed to the UDW detector undergoing a thermalization process after interacting with the environment. {In the case of relatively low temperatures, the difference of uncertainty and quantum coherence are mainly caused by different $\zeta$, and the larger $\zeta$ will induce the larger coherence and the smaller uncertainty.} High temperatures reduce the impact of boundary conditions on uncertainty, as shown in Figs. \ref{fig.5} (a-c) and (g-i). Theoretically, high temperatures cause the boundary condition related term (${k\zeta \sin \frac{{\Omega {\rm csch}\frac{{2\pi T}}{k}}}{{\pi T}}}$) of ${\mu _2}$ to approach zero,  thereby eliminating the effects of the different boundary conditions.

{In AdS spacetime, when the system's initial state is in a superposition, at low temperatures, the uncertainty and quantum coherence exhibit oscillatory behavior in the early evolution as well, while high temperatures suppress oscillations. The reasons for the generation of oscillations and the suppression at high temperatures are similar to those in dS spacetime. The oscillations arise due to the oscillatory terms $\cos\tilde \Omega \tau$ in $m_2$, while the suppression is caused by thermal noise.}

{ As mentioned earlier, from Figs. \ref{fig.5} and  \ref{fig.6}, it is easy to notice that the uncertainty and quantum coherence detected in the AdS  spacetime may differ due to different boundary conditions in early evolution, whereas, both of them will eventually converge to the constant values ${U}_s^{AdS}$ and ${C}_s^{AdS}$. To further analyze the factors influencing this constant value, we examine the expression for ${U}_s^{AdS}$ and ${C}_s^{AdS}$ when $\tau  \to \infty $. It} is interesting to find that ${U}_s^{AdS}$ and ${C}_s^{AdS}$ detected in Ads spacetime are identical  as those in dS spacetime, namely,
\begin{align}
{{U}_s^{AdS}(\tau  \to\infty)={U}_s^{dS}(\tau  \to\infty)=U_s,} \\
{{C}_s^{AdS}(\tau  \to\infty)={C}_s^{ds}(\tau  \to\infty)=C_s.}
\end{align}
That is to say, in AdS spacetime, the eventual evolutions of the observed uncertainty and coherence are also only related to the ratio $\Omega/T$. One can obtain lower uncertainty and more quantum coherence in the case of higher $\Omega/T$. This finding may provide the superiority for achieving quantum information processing tasks in AdS spacetime.

\section{discussions and Conclusions}
{We have investigated the ability of the Unruh-DeWitt detector to probe quantum uncertainty and quantum coherence in both de Sitter and Anti-de Sitter spacetimes. By analyzing the response functions of different UDW detectors, we have successfully identified the key factors affecting quantum uncertainty and quantum coherence in these curved spacetime backgrounds. Several interesting results are obtained as:

(i) In both dS and AdS spacetimes, the UDW detector eventually thermalizes with the environment due to long-term interactions, leading to the constant values of the uncertainty and quantum coherence.  It is interesting to find that  the constant values of the final uncertainty and coherence are identical as those in dS and AdS spacetimes, which are only related to the ratio of the energy gap $\Omega$ to the temperature $T$ in both spacetimes. To be explicit, the larger the ratio $\Omega/T$, the greater constant value of quantum coherence and the smaller   constant value of  uncertainty. Furthermore,  it reveals that the quantum uncertainty and quantum coherence exhibit a negative correlation in both spacetimes.

(ii) It turns out that the effect of temperature on quantumness in both dS and AdS spacetimes   is  prominent. Basically, the temperature $T$ is closely related to thermal noise, and an increasing $T$ will lead to enhanced thermal noise, which disrupts quantum coherence and consequently increases the system's uncertainty. When the system is in an initial superposition state, both uncertainty and quantum correlations exhibit oscillatory behavior at lower temperatures in both spacetimes. However,   this oscillatory behavior can be suppressed for higher $T$, while the increasing energy gap enhances the oscillations.

(iii) It is revealed that the acceleration's impact on uncertainty and quantum coherence in dS spacetime is significant during the early stages of evolution, while this acceleration does not alter the constant values of the final $U$ and $C$. Furthermore, it is observed that as the temperature increases, the influence of acceleration on uncertainty gradually weakens until it disappears. Besides, the boundary conditions of AdS spacetime have a similar effect on uncertainty and coherence in dS spacetime.

(iv)}  There are quite different sensitivities of the UDW detectors for temperature changes in these two spacetimes. In AdS spacetime, the UDW detectors of interest are more sensitive to temperature than those in dS spacetime. This means the UDW detector in AdS spacetime has already thermalized under the same temperature conditions, thereby eliminating the influence of other factors such as different boundary conditions on uncertainty or coherence.

With all these in mind, we argue that the current observations can offer  insights into the quantumness in de Sitter and Anti-de Sitter spacetimes, {and are beneficial to achieve prospective quantum tasks in the frame of curved spacetimes.}

\section*{Acknowledgements} %致谢（项目支持）
This work was supported by the National Science Foundation of China under (Grant Nos. 12475009, and 12075001), Anhui Provincial Key Research and Development Plan (Grant No. 2022b13020004), Anhui Province Science and Technology Innovation Project (Grant No. 202423r06050004), and Anhui Provincial University Scientific Research Major Project (Grant No. 2024AH040008).

\newcommand{\PRL}{\emph{Phys. Rev. Lett.} }
\newcommand{\RMP}{\emph{Rev. Mod. Phys.} }
\newcommand{\PRA}{\emph{Phys. Rev. A} }
\newcommand{\PRB}{\emph{Phys. Rev. B} }
\newcommand{\PRE}{\emph{Phys. Rev. E} }
\newcommand{\PRD}{\emph{Phys. Rev. D} }
\newcommand{\APL}{\emph{Appl. Phys. Lett.} }
\newcommand{\NJP}{\emph{New J. Phys.} }
\newcommand{\JPA}{\emph{J. Phys. A} }
\newcommand{\JPB}{\emph{J. Phys. B} }
\newcommand{\OC}{\emph{Opt. Commun.} }
\newcommand{\PLA}{\emph{Phys. Lett. A} }
\newcommand{\EPJD}{\emph{Eur. Phys. J. D} }
\newcommand{\NP}{\emph{Nat. Phys.} }
\newcommand{\NC}{\emph{Nat. Commun.} }
\newcommand{\EPL}{\emph{Europhys. Lett.} }
\newcommand{\AoP}{\emph{Ann. Phys.} }
\newcommand{\ADP}{\emph{Ann. Phys. (Berlin)} }
\newcommand{\QIC}{\emph{Quantum Inf. Comput.} }
\newcommand{\QIP}{\emph{Quantum Inf. Process.} }
\newcommand{\CPB}{\emph{Chin. Phys. B} }
\newcommand{\IJTP}{\emph{Int. J. Theor. Phys.} }
\newcommand{\IJMPB}{\emph{Int. J. Mod. Phys. B} }
\newcommand{\PR}{\emph{Phys. Rep.} }
\newcommand{\SR}{\emph{Sci. Rep.} }
\newcommand{\LPL}{\emph{Laser Phys. Lett.} }
\newcommand{\OEE}{\emph{Opt. Exp.} }
\newcommand{\IJQI}{\emph{Int. J. Quantum Inf.} }
\newcommand{\EPJC}{\emph{Eur. Phys. J. C} }

\bibliographystyle{plain}

\end{document}